\documentclass[useAMS,usenatbib]{aa}

\usepackage{amssymb}
\usepackage[fleqn]{amsmath}

\usepackage{times}
\usepackage{graphicx}
\usepackage{epstopdf}
\usepackage[T1]{fontenc}
\usepackage{aecompl} 
\usepackage{multirow}
\usepackage{pdflscape}
\usepackage{rotating}
\usepackage{float}

\usepackage{anyfontsize}

\usepackage{txfonts}
\usepackage{url}
\usepackage[colorlinks=true,linkcolor=blue,anchorcolor=blue,citecolor=blue,urlcolor=blue]{hyperref}

\usepackage{cleveref}

\newcommand{\beq}{\begin{equation}}
\newcommand{\eeq}{\end{equation}}

\def\gs{\mathrel{\lower0.6ex\hbox{$\buildrel {\textstyle >}\over{\scriptstyle \sim}$}}}
\def\ls{\mathrel{\lower0.6ex\hbox{$\buildrel {\textstyle <}\over{\scriptstyle \sim}$}}}
\newcommand{\simgt}{\lower.5ex\hbox{$\; \buildrel > \over \sim \;$}}
\newcommand{\simlt}{\lower.5ex\hbox{$\; \buildrel < \over \sim \;$}}

\defcitealias{se+et15_comalit_I}{CoMaLit-I}
\defcitealias{ser+al15_comalit_II}{CoMaLit-II}
\defcitealias{ser15_comalit_III}{CoMaLit-III}
\defcitealias{se+et15_comalit_IV}{CoMaLit-IV}
\defcitealias{se+et17_comalit_V}{CoMaLit-V}
\defcitealias{xxl_I_pie+al16}{XXL~Paper~I}
\defcitealias{xxl_II_pac+al16}{XXL~Paper~II}
\defcitealias{xxl_III_gil+al16}{XXL~Paper~III}
\defcitealias{xxl_IV_lie+al16}{XXL~Paper~IV}
\defcitealias{xxl_XIII_eck+al16}{XXL~Paper~XIII}
\defcitealias{xxl_XX_ada+al18}{XXL~Paper~XX}
\defcitealias{xxl_XXV_pac+al18}{XXL~Paper~XXV}
\defcitealias{xxl_XVI_mar+al18}{XXL~Paper~XVI}

\begin{document}

\title{The XXL Survey\thanks{Based on observations obtained with XMM-Newton, an ESA science mission with instruments and contributions directly funded by ESA Member States and NASA. Based on observations made with ESO Telescopes at the La Silla and Paranal Observatories under programme ID 089.A-0666 and LP191.A-0268.}}
\subtitle{XXXVIII. Scatters and correlations of X-ray proxies in the bright XXL cluster sample}

\author{Mauro Sereno \inst{1,2}\fnmsep\thanks{\email{\href{mailto:mauro.sereno@inaf.it}{mauro.sereno@inaf.it}}}
          \and
          Stefano Ettori\inst{1,2}
          \and
          Dominique Eckert\inst{3}
          \and
          Paul Giles\inst{4,5}
          \and
          Ben J. Maughan\inst{5}
          \and
          Florian Pacaud\inst{6}
          \and
          Marguerite Pierre\inst{7}
          \and 
          Patrick Valageas\inst{8} 
          }

\institute{
	INAF - Osservatorio di Astrofisica e Scienza dello Spazio di Bologna, via Piero Gobetti 93/3, I-40129 Bologna, Italy
	\and INFN, Sezione di Bologna, viale Berti Pichat 6/2, I-40127 Bologna, Italy 
	\and Department of Astronomy, University of Geneva,   ch. d'Ecogia 16, CH-1290 Versoix, Switzerland
	\and Astronomy Centre, University of Sussex, Falmer, Brighton, BN1 9QH, UK
	\and H. H. Wills Physics Laboratory, University of Bristol, Tyndall Ave, Bristol BS8 1TL, UK
	\and Argelander Institut f\"ur Astronomie,  Universit\"at Bonn, D-53121 Bonn, Germany
	\and AIM, CEA, CNRS, Universit\'e Paris-Saclay, Universit\'e Paris Diderot, Sorbonne Paris Cit\'e, F-91191 Gif-sur-Yvette, France
	\and Institut de Physique Theorique, CEA, Saclay, F-91191 Gif sur Yvette, France
	}

\abstract
{Scaling relations between cluster properties embody the formation and evolution of cosmic structure. Intrinsic scatters and correlations between X-ray properties are determined from merger history, baryonic processes, and dynamical state.} 
 {We look for an unbiased measurement of the scatter covariance matrix between the three main X-ray observable quantities attainable in large X-ray surveys --  temperature, luminosity, and gas mass. This also gives us the cluster property with the lowest conditional intrinsic scatter at fixed mass.}
{Intrinsic scatters and correlations can be measured under the assumption that the observable properties of the intra-cluster medium hosted in clusters are log-normally distributed around power-law scaling relations. The proposed method is self-consistent, based on minimal assumptions, and requires neither the external calibration by weak lensing, dynamical, or hydrostatic masses nor the knowledge of the mass completeness.}
{We analyzed the 100 brightest clusters detected in the XXL Survey and their X-ray properties measured within a fixed radius of 300~kpc. The gas mass is the less scattered proxy ($\sim8\%$). The temperature ($\sim20\%$) is intrinsically less scattered than the luminosity ($\sim30\%$) but it is measured with a larger observational uncertainty. We found some evidence that gas mass, temperature and luminosity are positively correlated. Time-evolutions are in agreement with the self-similar scenario, but the luminosity-temperature and the gas mass-temperature relations are steeper.}
{Positive correlations between X-ray properties can be determined by the dynamical state and the merger history of the halos. The slopes of the scaling relations are affected by radiative processes.}
 
\keywords{
surveys, X-rays: general, X-rays: galaxies: clusters, galaxies: clusters: intracluster medium,  cosmology: large-scale structure
               }

\authorrunning{M. Sereno et al.}
\titlerunning{Cluster X-ray proxies}
\maketitle

\section{Introduction}

The physics of baryons and dark matter can be assessed with scaling relations between cluster properties  \citep{pra+al09,arn+al10,gio+al13}. Ongoing and upcoming large surveys are measuring a wealth of cluster properties, e.g. optical richness, X-ray luminosity, Sunyaev-Zel'dovich (SZ) flux \citep{eucl_lau_11,planck_2013_XXIX,ble+al15,xxl_I_pie+al16,mel+al17,mat+al19}. 

Gravity is the driving force in structure formation and evolution, and makes clusters self-similar with observable properties following power-law relations in halo mass \citep{kai86,gio+al13,ett13}. Deviations from the self-similar scheme are due to non-gravitational processes, such as feedback and non-thermal mechanisms, which can contribute significantly to the global energy budget \citep{mau+al12}. 

The scaling relations are scattered by underlying processes that can affect different cluster properties to different degrees \citep{sta+al10,tru+al18}. Numerical simulations \citep{sta+al10,fab+al11,ang+al12,sar+al13} and observational studies \citep{mau07,vik+al09} confirm that the properties are log-normally distributed. Broadly speaking, the scatter is related to the regularity of the clusters \citep{se+et15_comalit_I,se+et15_comalit_IV} and to deviations from equilibrium \citep{fab+al11,sar+al13}. Well behaved proxies with small scatters can be used to provide accurate measurement of the mass of galaxy clusters, which is crucial in important branches of cosmology and astrophysics \citep{ett+al09,vik+al09b,man+al10,planck_2013_XX}. 

If we measure scalings, scatters, and correlations of scaling relations, we can study the forces driving cluster formation and evolution. In this paper, we investigate X-ray properties measurable in large surveys. We propose a novel statistical method where scatters are measured exploiting the expected linearity of the relations even without knowing the mass of the clusters. An observable property of a galaxy cluster can be a good proxy if it is easy to measure and well behaved. Intrinsic scatter enables us to determine such a variable. We can view the proxy with the lowest intrinsic conditional scatter with respect to some basic cluster characteristic as `optimal'. 

We exploit the XXL Survey, the largest completed XMM-Newton project \citep[\citetalias{xxl_I_pie+al16}]{xxl_I_pie+al16}.  Two sky regions for a total of 50 square degrees has been surveyed and several hundreds of galaxy clusters out to redshift $\sim2$ have been detected \citep[\citetalias{xxl_XX_ada+al18}]{xxl_XX_ada+al18}. We consider the three main X-ray global quantities measured by the mission, i.e. temperature, luminosity, and gas mass.

The paper is as follows. The regression method and its extension to multi-response variables are described in Secs.~\ref{sec_regr} and \ref{sec_mult}, respectively. The data sample is introduced in Sec.~\ref{sec_samp}. Theoretical expectations are briefly discussed in Sec.~\ref{sec_theo}. Section~\ref{sec_resu} presents the results. Section~\ref{sec_prev} is devoted to comparison with previous analyses. Final considerations are in Sec.~\ref{sec_conc}. Appendix~\ref{app_asym} presents a simple recipe to deal with asymmetric errors and logarithmic variables. Additional figures are presented in App.~\ref{app_2D}.



\subsection{Notations}

The frame-work cosmological model in use in the XXL papers is the flat $\Lambda$CDM universe with density parameter $\Omega_\text{M}=0.28$, and Hubble constant $H_0=70~\text{km~s}^{-1}\text{Mpc}^{-1}$, as found from the study of the final nine years cosmic microwave background observations of the Wilkinson Microwave Anisotropy Probe satellite (WMAP9), combined with baryon acoustic oscillation measurements and constraints on $H_0$ from Cepheids and type Ia supernovae \citep{hin+al13}. 

As usual, $H(z)$ is the redshift dependent Hubble parameter and $E_z\equiv H(z)/H_0$. When $H_0$ is not specified, $h$ is the Hubble constant in units of $100~\mathrm{km~s}^{-1}\mathrm{Mpc}^{-1}$. 

$O_{\Delta}$ denotes a cluster property measured within the radius $r_{\Delta}$ which encloses a mean over-density of $\Delta$ times the critical density at the cluster redshift, $\rho_\mathrm{cr}=3H(z)^2/(8\pi G)$. 

`$\log$' is the logarithm to base 10 and `$\ln$' is the natural logarithm. Results for natural logarithm are quoted as percents, i.e. 100 times the dispersion in natural logarithm. 

By intrinsic scatter we mean the standard deviation of the conditional probability, e.g., the probability of the temperature given the mass. If the conditional probability is related to the measurement process, i.e. the probability of the measured output given true input, we name it statistical measurement uncertainty. We do not name as scatter the standard deviation of the marginalized distribution, e.g. the distribution of temperatures. Throughout the paper, we denote the intrinsic scatter as $\sigma$ and the measurement uncertainty as $\delta$.

\section{Regression scheme}
\label{sec_regr}

\begin{figure}
\resizebox{\hsize}{!}{\includegraphics{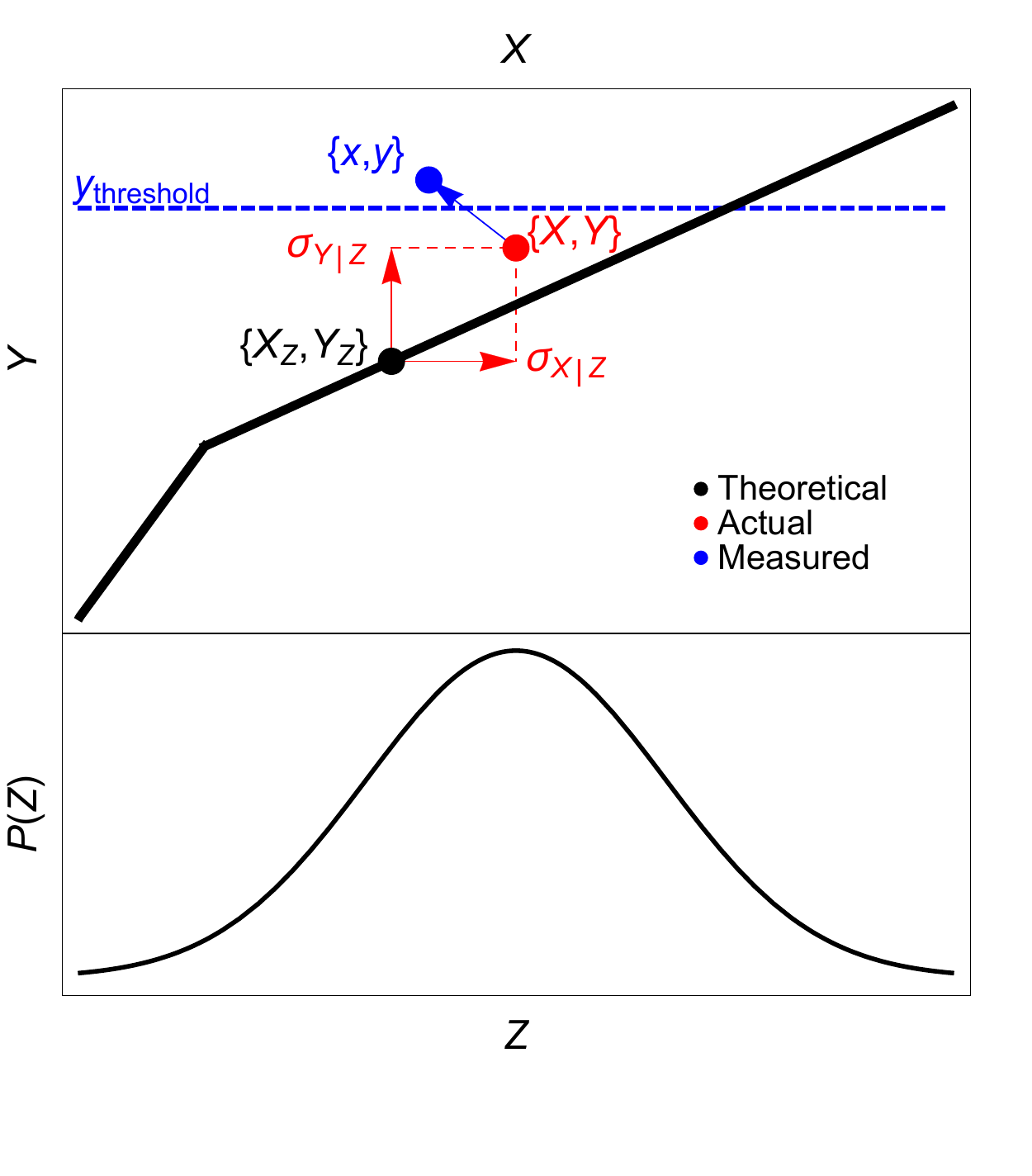}}
\caption{Graphic view of some quantities playing in the Bayesian regression scheme, see Sec.~\ref{sec_regr}: the measurement results (blue); the true property values (red); the rescaled values of a basic feature (black).
}
\label{fig_regression}
\end{figure}

In this section, we describe the Bayesian fitting procedure used to retrieve the scaling relations and the intrinsic scatters. We assume that the cluster properties (temperature, luminosity, gas mass) are power laws of some basic property, e.g. the mass. Hereafter, we focus on the logarithms of these quantities, which are thus linearly related with each other.

In a nutshell, we first appoint a basic intrinsic cluster feature (denoted by $Z$ as the reasoning would apply to the mass as well as to other choices). For any measurable property, e.g. the temperature, we distinguish three variables, see Fig.~\ref{fig_regression}: i) $X_Z$, the latent quantity that is exactly linked to $Z$ through a functional relation $X(Z)$ \citep{mau14}. ii) $X$, the true quantity that would be measured in an theoretical observation with infinite accuracy and precision \citep{fe+ba12}.  $X$ is intrinsically scattered with respect to $X_Z$. iii) $x$, the manifest result of the measurement process which shows some observational noise with respect to $X$.

Similarly to $X$, we can consider an additional cluster property $Y$ and the related $y$ and $Y_Z$. If the latent variables ($X_Z$, $Y_Z$) are linearly related to $Z$, they are linearly related with each other. We can then identify the best proxy with the X-ray observable characterized by the lowest conditional intrinsic scatter with respect to $Z$. In this way, the optimal proxy can be established even if we do not know the cluster mass. 

Of course, we still need the mass if we want to calibrate the scaling relations and unambiguously identify the best proxy as the best mass proxy. In principle, $Z$ can be any fundamental property of the cluster. Since we do not measure the mass itself, our analysis is based on the fact that $Z$ is the quantity that can best characterize the X-ray properties of the clusters and minimize the scatters. Even though on a theoretical basis the most suitable candidate for this role is the mass, it is not guaranteed that $Z$ is the mass. It could be another quantity, e.g. the optical richness or the SZ signal, or one of the three X-ray properties measured in the present paper. However, this last hypothesis can be discarded if the estimated intrinsic scatters are not null.

\subsection{Scaling and distributions}

In order to quantify the intrinsic scatter of the X-ray properties, we follow the regression scheme detailed in the CoMaLit series \citep[Comparing Masses in Literature,][]{se+et15_comalit_I,ser+al15_comalit_II,se+et15_comalit_IV,se+et17_comalit_V} and in \citet{ser16_lira}. This scheme accounts for time-evolution, correlated intrinsic scatters, and selection effects (Malmquist/Eddington biases). 

In this section we consider a pair of observables. In the next section, we generalize the procedure to the case of multiple response variables. 

As result of the measurement of the $j$-th cluster, the observable $x_j$, $y_j$, and the related uncertainty covariance matrix  $\mathbf{V}_{\delta,j}$ are known. On the other hand, $\{X_{Z,j},Y_{Z,j}\}$, $\{X_j,Y_j\}$, and the covariance matrix of the intrinsic scatters, $\mathbf{V}_{\sigma,j}$, are unknowns to be determined under the assumption of linearity. 

In case of a linear relation, $X_Z$ and $Y_Z$ are related to the same covariate variable, $Z$, as
\begin{align}
X_Z & =  \alpha_{X|Z}+\beta_{X|Z} Z + \gamma_{X|Z} \log F_z, \label{eq_bug_1} \\
Y_Z & =  \alpha_{Y|Z}+\beta_{Y|Z} Z + \gamma_{Y|Z} \log F_z, \label{eq_bug_2}
\end{align}
where $\alpha$ denotes the normalization, the slope $\beta$ accounts for the dependence with $Z$, and the slope $\gamma$ accounts for the redshift-evolution. $F_z$ is the re-normalized Hubble parameter, $F_z=E_z/E_z(z_\text{ref})$. Here and in the following, we assume a power-law dependence on $F_z$ for the redshift-evolution of the observables. 

The relations between $Z$, $X_Z$, and $Y_Z$ are deterministic and they are not affected by scatter. We assume that the uncertainty on the spectroscopic redshift $z$ is negligible. 

If we do not know the value of $Z$, the two slopes and the two normalizations in Eqs.~(\ref{eq_bug_1}) and (\ref{eq_bug_2}) are redundant. We then assume that $X$ is an unbiased proxy of $Z$, i.e. we fix $\alpha_{X|Z}=0$, $\beta_{X|Z}=1$, and $\gamma_{X|Z}=0$. Fixing the parameters of the $X$-$Z$ rather than the $Y$-$Z$ relation is just a matter of rescaling which does not affect the analysis of the intrinsic scatters. In absence of a direct measurement of $Z$, any bias between $X$ and $Z$ (i.e. $\alpha_{X|Z}\neq0$) is degenerate with the estimated overall normalization of the scaling between $Y$ and $Z$. The data analysis can only constrain the relative bias between $X$ and $Y$ \citep[CoMaLit-I]{se+et15_comalit_I}. 

The measured and the true values of the quantities are related as
\beq
P(x_j,y_j | X_j,Y_j ) \propto {\cal N}^\text{(2)}(\{X_j,Y_j\},\mathbf{V}_{\delta,j}) {\cal U}(y_{\text{th},j},\infty), \label{eq_bug_6}
\eeq
where ${\cal N}^\text{(2)}$ is the bivariate Gaussian distribution and ${\cal U}$ is the uniform distribution. $\mathbf{V}_{\delta,j}$ is the uncertainty covariance matrix whose diagonal elements are denoted as $\delta_{x,j}^2$ and $\delta_{y,j}^2$, and whose off-diagonal elements are denoted as $\rho_{xy,j}\delta_{x,j}\delta_{y,j}$. The proportionality symbol in Eq.~(\ref{eq_bug_6}) indicates that the function on the right hand side is not normalized.

The probability distribution in Eq.~(\ref{eq_bug_6}) is truncated for $y_j<y_{\text{th},j}$, which accounts for selection effects when only clusters above an observational threshold (in the response variable) are included in the sample, i.e. the Malmquist bias \citep[CoMaLit-II]{ser+al15_comalit_II}. 

To shorten the notation in Eq.~(\ref{eq_bug_6}) and similarly in the following, we drop the explicit dependence on the fixed parameters, i.e. $\mathbf{V}_{\delta,j}$ and $y_{\text{th},j}$, on the left hand side. On the right hand side, we do not express the functional dependence on the random variables $x_j$ and $y_j$.

The observational threshold $y_\text{th}$ may not be exactly known. This is the case when the quantity which the selection procedure is based on differs from the quantity used in the regression. As an example, XXL clusters are selected according to their flux within 1\arcmin\ whereas we consider the luminosity within 0.3~Mpc in the regression, see Sec.~\ref{sec_samp}. We have then to consider the additional relation
\beq
P(y_{\text{th},j}|y_{\text{th,obs},j}) = {\cal N}(y_{\text{th,obs},j},\delta_{y_{\text{th},j}}^2) \label{eq_mb_1},
\eeq
where $\delta_{y_{\text{th},j}}$ is the uncertainty associated to the measured threshold $y_{\text{th,obs},j}$. This save us from adding new quantities, e.g. the observed flux, to the regression scheme. 

We assume that the intrinsic scatters are Gaussian. It is
\beq
P(X_j,Y_j | X_{Z,j},Y_{Z,j} )  \propto  {\cal N}^\text{(2)}(\{X_{Z,j},Y_{Z,j}\},\mathbf{V}_{\sigma,j}) {\cal U}(Y_{\text{th},j},\infty), \label{eq_bug_3}
\eeq
where $\mathbf{V}_{\sigma,j}$ is the scatter covariance matrix whose diagonal elements are denoted as $\sigma_{X|Z,j}^2$ and $\sigma_{Y|Z,j}^2$, and whose off-diagonal elements are denoted as $\rho_{XY|Z,j}\sigma_{X|Z,j}\sigma_{Y|Z,j}$. 

$Y_{\text{th},j}$ is the threshold in the response variable,
\beq
P(Y_{\text{th},j}|y_{\text{th},j})={\cal N}(y_{\text{th},j},\delta_{y,j}^2). \label{eq_mb_2}
\eeq
Even though the selection procedure is based only on the value of the measured $y$ rather than the value of $Y$, any threshold in $y$ affects all the probability distributions. In fact, we do not sample a generic distribution of clusters but we select them and we have to model the distribution of the sampled objects. Hence, the distribution of $Y$ given $Z$ for a generic cluster from the full population differs from the distribution specific to a selected sample, which follows Eq.~(\ref{eq_bug_3}) and it is truncated. 

We assume that the scatters are time-dependent, hence the subscript $j$ in the covariance matrix, but do not depend on $Z$ \citep{roz+al10}. The time evolution of the intrinsic scatters and of the correlation can be modeled as
\begin{align}
\sigma_{X|Z}(z) & = \sigma_{X|Z,0}F_z^{\gamma_{\sigma_{X|Z}}}, \label{eq_bug_4} \\
\sigma_{Y|Z}(z) & =  \sigma_{Y|Z,0}F_z^{\gamma_{\sigma_{Y|Z}}}, \label{eq_bug_5} \\
\rho_{XY|Z}(z) & =  \rho_{XY|Z,0}F_z^{\gamma_{\rho_{XY|Z}}}. \label{eq_bug_5b} 
\end{align}

The intrinsic distribution of $Z$ can be approximated with a mixture of Gaussian functions  (\citealt{kel07}, \citetalias{ser+al15_comalit_II}, \citealt[CoMaLit-IV]{se+et15_comalit_IV}). In the simplest but still effective case of one component \citep{ser16_lira},
\beq
P(Z) = {\cal N}\left(\mu_Z, \sigma_{Z}^2 \right). \label{eq_bug_7}
\eeq
Most of the parent populations of astronomical quantities, e.g. the halo mass function or the luminosity function, are locally exponential (in log-space), i.e. $P_\text{parent} (Z) \sim \exp (-a Z)$. However, here we have to model just the distribution of the clusters included in the sample rather than the full population. Once the parent population is filtered by the selection process, a Gaussian distribution provides a reliable approximation \citepalias{se+et15_comalit_IV}.

The redshift evolution of the (mean of the)  $Z$-distribution can be modeled as \citepalias{se+et15_comalit_IV},
\beq
\label{eq_bug_8}
\mu_Z (z) = \mu_{Z,0} +\gamma_{\mu_Z}\log F_z + \gamma_{\mu_Z,D}\log D_z,
\eeq
where $\mu_{Z,0}$ is the local mean and $D_z$ is the luminosity distance. We renormalize the distances such that $D_z(z_\text{ref})=1$.

The dispersion of the $Z$-distribution evolves as
\beq
\label{eq_bug_9}
\sigma_{Z}(z)=\sigma_{Z,0}F_z^{\gamma_{\sigma_{Z}}}.
\eeq

The dependence on $F_z$ is enough to account for the redshift evolution of the scaling relations, see Eqs.~(\ref{eq_bug_1},~\ref{eq_bug_2}) and of the scatters, see Eqs.~(\ref{eq_bug_4}--\ref{eq_bug_5b}) and Eq.~(\ref{eq_bug_9}). This is justified by theoretical predictions based on the self-similar model, by results of numerical simulations, and by observational fits \citepalias{se+et15_comalit_IV}. On the other hand, we introduce an explicit dependence on the cosmological distance for the evolution of the covariate distribution, see Eq.~(\ref{eq_bug_8}). In fact, the completeness of a sample selected according to either flux or signal-to-noise depends on the distance \citepalias{se+et15_comalit_IV}. The redshift dependence in Eq.~(\ref{eq_bug_8}) is general enough to address even more complicated cases.

\subsection{Priors}
\label{sec_regr_priors}

Priors have to be conveniently non-informative \citepalias{se+et15_comalit_I,ser+al15_comalit_II}. The priors on the intercept $\alpha_{Y|Z}$ and on the mean $\mu_{Z,0}$ are flat,
\beq
\label{eq_bug_10}
\alpha_{Y|Z},\  \mu_{Z,0}  \sim  {\cal U}(-1/\epsilon,1/\epsilon),
\eeq
where $\epsilon$ is a small number. In our calculation, $\epsilon = 10^{-4}$. 
The prior on the correlation $\rho_{XY|Z,0}$ is flat too,
\beq
\label{eq_bug_11}
\rho_{XY|Z,0} \sim  {\cal U}(-1,1).
\eeq

We model the prior probability of the slopes and of the time-evolution as a Student's $t_1$ distribution with one degree of freedom,
\beq
\label{eq_bug_12}
\beta_{Y|Z},\ \gamma_{Y|Z},\ \gamma_{\mu_Z},\ \gamma_{\mu_Z,D} \sim  t_1.
\eeq
This is equivalent to uniformly distributed direction angles $\arctan \beta$ and $\arctan \gamma$.

As non informative priors for the evolution of scatters and correlations, we consider uniform distributions. Since intrinsic scatters are expected to slightly increases with redshifts \citepalias{se+et15_comalit_IV}, we assume
\beq
\label{eq_bug_12b}
\gamma_{\sigma_{X|Z}},\ \gamma_{\sigma_{Y|Z}} \sim  {\cal U}(0,1),
\eeq
and
\beq
\label{eq_bug_12c}
\gamma_{\rho_{XY|Z}} \sim  {\cal U}(-1,1).
\eeq
For the evolution of the dispersion of the $Z$-distribution we adopt
\beq
\label{eq_bug_12d}
\gamma_{\sigma_Z} \sim  {\cal U}(-1,1).
\eeq
For the precision, i.e. the inverse of the variance, we adopt a nearly scale-invariant Gamma distribution,
\beq
\label{eq_bug_13}
1/\sigma_{X|Z,0}^2,\ 1/\sigma_{Y|Z,0}^2,\ 1/\sigma_{Z,0}^2 \sim \Gamma(r,\lambda),
\eeq
where the rate $r$ and the shape parameter $\lambda$ are fixed to $r=\lambda=\epsilon$ so that the prior spans a considerable range and it is nearly constant in logarithmic bins.

Alternatively to non-informative priors, we consider strong assumptions. The evolution of scatters and correlations is poorly constrained in samples of $\sim100$ objects \citep{ser16_lira}. In our reference regression scheme, we only constrain the redshift weighted scatters and we fix
\beq
\label{eq_bug_12a}
\gamma_{\sigma_{X|Z}} = \gamma_{\sigma_{Y|Z}}=\gamma_{\rho_{XY|Z}} = 0 .
\eeq

\section{Multi-response regression scheme}
\label{sec_mult}

The scheme detailed in Section~\ref{sec_regr} can be generalized to the simultaneous regression of $n(\ge2)$ observables. In this scheme, $y_{ij}$ is the $j$-th measurement of the $i$-th observable, $Y_{ij}$ is the true value, and $Y_{Z,ij}$ is the latent unscattered quantity which fits the scaling relation. To simplify the notation, in the following we dismiss the subscripts for the time dependence of the scatters. The scaling relation of the $i$-th property is expressed as
\beq
\label{eq_bug_multi_1}
Y_{Z,i}  =  \alpha_{Y_i|Z}+\beta_{Y_i|Z} Z + \gamma_{Y_i|Z} \log F_z .
\eeq
Due to degeneracy, we anchor the scaling parameters of the first response variable, i.e. $\alpha_{Y_1|Z}=0$, $\beta_{Y_1|Z}=1$, and $\gamma_{Y_1|Z}=0$. The reference $Z$ variable is modeled as in Eq.~(\ref{eq_bug_7}). In absence of Malmquist biases, $Z$ is the expected values of $Y_1$given $Z$, $\langle Y_1|Z \rangle =Z$.

The intrinsic scatters shape the distribution of the true quantities around the model predictions. For the $j$-th cluster
\begin{align}
P( Y_{1,j},Y_{2,j},.... |  Y_{Z,1j},Y_{Z,2j},....) \propto {\cal N}^{{(n)}}\left(\{Y_{Z,1j},Y_{Z,2j},....\},\mathbf{V}_{\sigma}\right)  \label{eq_bug_multi_3} \\
 \times  \prod_i {\cal U}(Y_{\text{th},ij},\infty),  \nonumber
\end{align}
where ${\cal N}^{(n)}$ is the multivariate Gaussian distribution. $\mathbf{V}_{\sigma}$ is the $n\times n$ scatter covariance matrix whose diagonal elements are the intrinsic variances, $\sigma_{Y_i|Z}^2$, and whose off-diagonal elements can be expressed in terms of the correlations as $\rho_{Y_aY_b|Z}\sigma_{Y_a|Z}\sigma_{Y_b|Z}$. The threshold for the $j$-th measurement of the $i$-th response variable is $Y_{\text{th},ij}$.

The measured and the true values are related as
\begin{align}
P( y_{1,j},y_{2,j},... | Y_{1,j},Y_{2,j},...) & \propto \ {\cal N}^{(n)}\left(\{Y_{1,j},Y_{2,j}\},\mathbf{V}_{\delta,j}\right) \label{eq_bug_multi_2}\\
& \times  \prod_i {\cal U}(y_{\text{th},ij},\infty), \nonumber
\end{align}
where $\mathbf{V}_{\delta,j}$ is the $n\times n$ uncertainty covariance matrix of the $j$-th cluster. The thresholds for the measured and the true response values are related as
\beq
\label{eq_bug_multi_4}
P(y_{\text{th}ij}|y_{\text{th,obs,}ij})={\cal N}\left(y_{\text{th,obs,}ij},\delta_{y_\text{th,ij}}^2\right),
\eeq
and
\beq
\label{eq_bug_multi_5}
P(Y_\text{th,ij}|y_\text{th,ij})={\cal N}\left(y_\text{th,ij},\delta_{y,ij}^2\right),
\eeq
where $y_{\text{th,obs,}ij}$ is the observational threshold, $\delta_{y_\text{th,ij}}$ is the related uncertainty, and $\delta_{y,ij}$ is the uncertainty associated to $y_{ij}$.

Similarly to Sec.~\ref{sec_regr}, Malmquist bias is treated with the inclusion of the (smooth) truncations in Eqs.~(\ref{eq_bug_multi_3},~\ref{eq_bug_multi_2}). 

\subsection{Priors}

We express the prior on the (inverse of the) intrinsic scatter matrix in terms of the Wishart distribution,
\beq
\mathbf{V}_{\sigma}^{-1} \sim \mathbf{W}(\mathbf{S},d),
\eeq
where $d$ is the number of degrees of freedom and $\mathbf{S}$ in the $n\times n$ scale matrix. We take $d=n+1$, so that the marginalized prior distribution of the correlation factors is uniform between $-1$ and $1$. In analogy to the variances in Eq.~(\ref{eq_bug_13}), we model $\mathbf{S}$ as a scalar matrix with diagonal elements
\beq
\mathbf{S}_{aa} \sim \Gamma(\epsilon,\epsilon).
\eeq
The Wishart prior is widely regarded as non-informative, even though it favors high variance in case of high correlation.
Other priors are defined as in Sec.~\ref{sec_regr_priors}.

\section{The sample}
\label{sec_samp}

The XXL survey covers a total area of 50 square degrees with a X-ray sensitivity of $\sim 10^{-14}\mathrm{erg~cm^{-2}s^{-1}}$ in the [0.5-2] keV band for extended sources \citep[\citetalias{xxl_II_pac+al16}]{xxl_II_pac+al16}. The survey has uncovered more than 300 galaxy clusters out to redshift $\sim2$ \citepalias{xxl_XX_ada+al18} over a wide range of nearly two decades in mass \citep[\citetalias{xxl_IV_lie+al16}]{xxl_IV_lie+al16}. The XXL programme has the potential to constrain at the same time cluster scaling relations and cosmological parameters \citep[\citetalias{xxl_XXV_pac+al18}]{xxl_XXV_pac+al18}. 

We here consider the sample of the 100 brightest clusters (XXL-100-GC) from the first data release (DR1) \citepalias[DR1,][]{xxl_I_pie+al16}. The candidate clusters were selected by setting a lower limit of $3\times10^{-14}\mathrm{erg~cm^{-2}s^{-1}}$ on the source flux in the soft X-ray band within a 1\arcmin\ aperture \citepalias{xxl_II_pac+al16}. We analyse the  spectroscopic temperature within 300~kpc, $T_\mathrm{300kpc}$, the luminosity in the rest-frame soft band [0.5-2.0] keV within 300~kpc, $L^\mathrm{XXL}_\mathrm{300kpc}$ \citep[\citetalias{xxl_III_gil+al16}]{xxl_III_gil+al16} and the gas mass. Here, we consider the gas mass in a sphere of radius equal to 300~kpc, $M_\mathrm{gas,300kpc}$, which are computed following the procedure described in \citet[\citetalias{xxl_XIII_eck+al16}]{xxl_XIII_eck+al16}, which we refer to for details. The full set of measurements is available for 96 clusters. The sample has a mean redshift of $z=0.38$ with standard deviation of $\Delta z=0.24$. The sample covers an extended range in temperature, from small groups at $T_\mathrm{300kpc}\la 1~\text{keV}$ to rich clusters at $T_\mathrm{300kpc}\la 7~\text{keV}$, even though the most massive and rare clusters from the extreme tail of the cosmological halo mass function are absent due to the limited survey area coverage. For a comprehensive discussion of the sample properties we refer to \citetalias{xxl_II_pac+al16}.



In the following, we recover the intrinsic scatters and correlations of the X-ray properties without having access to the mass. We use a simplified notation in the log-space. The measured gas mass, temperature, and luminosity in logarithmic units are 
\begin{align}
m_\text{g} & =\log (M_\text{gas,300kpc}/10^{14}/M_\odot), \\
t & =\log (T_\text{300kpc}/\text{keV}), \\
l & =\log (L^\text{XXL}_\text{300kpc}/10^{44}/\text{erg}/\text{s}^{-1}).
\end{align}
For the latent variables, we cannot strictly follow the convention in Secs.~\ref{sec_regr} and \ref{sec_mult}, since X-ray observables in linear space are usually named by capital letters, e.g. $M_\text{gas,300kpc}$. Then, if $m_\text{g}$ is the measured gas mass ($y$), the hypothetical measurement in absence of noise ($Y$) is $m_\text{g}^{(Y)}$  and the unscattered gas mass ($Y_Z$) is $m_\text{g}^{(Z)}$. The same convention applies to temperature and luminosity.

For our analysis, we follow a data-driven approach and we consider as proxies only X-ray properties measured within fixed physical radii. The use of quantities measured in a given over-density radius as proxy can be ambiguous to some degree. If we know the over-density radius, we know by definition the mass too. Let $O_{\Delta}$ be a generic observable quantity, e.g. temperature or richness, within the over-density radius $r_\Delta$. In practice, we use only a part ($r_\Delta$) of the full information we already have (the mass $M_{\Delta}$) to get a deteriorated version, i.e. the mass proxy $M_{\Delta}(O_{\Delta})$ calculated through the scattered scaling relation applied to $O_{\Delta}$, of the main information itself (the mass $M_{\Delta}$). In this sense, we lose information. This can be corrected with iterative approaches by determining at the same time $r_\Delta$, $O_{\Delta}$, and $M_{\Delta}$. However, we have still to rely on very strong priors (usually, the knowledge of how $r_\Delta$ scales with some observable property). This has little effect if the observable is poorly correlated with the radius, e.g. the X-ray luminosity emitted from a very large area, but it is a major problem otherwise, e.g. the gas mass. Valuable mass proxies can be highly correlated with the integration radius. 

The use of X-ray properties measured within fixed physical radii also minimize the impact of the assumed cosmological parameters on our results \citep{se+et15_comalit_I}. A different frame-work cosmology would mostly imply a different normalization and a slightly different time-dependence of the relations. Slopes and scatters would be minimally affected. At present, analyses of number counts \citepalias{xxl_XXV_pac+al18}  and clustering \citep[\citetalias{xxl_XVI_mar+al18}]{xxl_XVI_mar+al18} of XXL clusters are compatible with our reference cosmological model.

\subsection{Covariance uncertainty matrix}
\label{sec_cova}

The knowledge of the covariance uncertainty matrix is crucial to obtain unbiased estimates of the intrinsic scatters and of their correlations. Measurements of luminosity and temperature are based on the spectroscopic analysis of the core region and are largely independent of the gas mass measurement process, which exploits the photometry and the surface brightness profile in annular regions. However, the photons are the same and the gas mass measurement process uses the estimated temperature to convert the observed surface-brightness profiles into emission-measure profiles. This conversion is largely insensitive to the temperature and metallicity as long as the temperature exceeds $\sim$1.5 keV. In most cases the temperature and the gas measurement are nearly uncorrelated.

To estimate the uncertainty covariance matrix we proceed in the following way. Luminosity and temperature are estimated in a single measurement process. Their correlation is an output of the spectroscopic analysis. We approximate the probability distribution of the observed luminosity and temperature as a bivariate Gaussian. 

Measured temperature and gas mass are correlated too. The estimate of the gas mass relies on the conversion of the observed surface brightness into the emission measure. The conversion factor is computed using $T_\mathrm{300kpc}$ and simulating a single-temperature absorbed thin-plasma model with the APEC code \citepalias{xxl_XIII_eck+al16}.

To estimate the full covariance uncertainty matrix, we extract $10^5$ couples of luminosity and temperature for each cluster from the approximated bivariate normal distribution. We then compute the new conversion factor from surface brightness into emission measure for the sampled temperature and we compute the corresponding gas mass by rescaling. We finally extract a new gas mass measurement from a normal distribution centered on the rescaled $M_\mathrm{gas,300kpc}$ and with standard deviation equal to the observational uncertainty $\delta M_\mathrm{gas,300kpc}$. The final correlation matrix is computed from the sampled values of temperature, luminosity, and gas mass.

\subsection{Selection effects}
\label{sec_sele}

Cosmological studies of number counts and abundance evolution require a very detailed study of the completeness. Observed properties have to be related to the underlying mass function and the selection function can be expressed in terms of the true cluster parameters rather than in terms of their measured counterparts affected by measurement errors. 

The selection and validation of the XXL-100-GC sample is described in \citetalias{xxl_II_pac+al16}. Here we just recall the main features relevant for our analysis. The XXL-100-GC sample was chosen with a flux-limit selection in a fixed angular aperture. The flux limit of the final catalog is $3\times10^{-14}\mathrm{erg~s^{-1}cm^{-2}}$ in the [0.5-2.0] keV band within a 1\arcmin\ radius aperture. Assuming temperature equal $3~\mathrm{keV}$, metallicity equal to $0.3$, and redshift $z=0.3$, this is equivalent to a total MOS1+MOS2+PN count rate of 0.0332~cts/s in [0.5-2] keV.

The C1+2 pipeline selection function of the XXL Survey depends on cluster profile, emissivity, and position \citepalias{xxl_II_pac+al16}. The dependence on the exposure and background level is encapsulated in the pointing under consideration and the off-axis angle is implicitly given by the sky position \citepalias{xxl_II_pac+al16}. Candidate clusters are then confirmed by visual inspection.

When we express the selection function in terms of the true intrinsic parameters, we have to consider that a cluster with a given true count rate can exceed the cut or fail depending on the flux measurement errors, which depend on the local exposure time and therefore the pointing on which the source was detected. Finally, in regions where several pointings overlap, the completeness must follow exactly the order and manner in which the two selection steps are applied to the source candidates.

Whereas the full information of the selection function and of the completeness of the sample is needed in cosmological studies, simpler methods are better suited to studies focused on the scaling relations. The CoMaLit approach can be applied to heterogeneous samples too. If we do not aim at expressing the observed distribution of clusters in terms of the underlying mass function, we have just to properly model the distribution to avoid Eddington/Malmquist biases. The distribution has to be flexible enough to fit the data. 

Within this framework, the Malmquist bias and the probability to exceed a given count rate can be simply modeled in terms of the observed flux as a step function rather than as a more complicated position dependent probability of the true flux.

The correction for the Malmquist bias is relevant in the case of the X-ray luminosity as response variable. The flux limit is set within a 1\arcmin\ aperture. We compute the luminosity threshold extrapolated out to 300~kpc at the cluster redshift, by assuming a $\beta$-profile with core radius $r_\text{c}=0.15\ r_{500}$ and $\beta=2
/3$ \citepalias{xxl_III_gil+al16}. This procedure gives the luminosity threshold $L^\mathrm{XXL}_\mathrm{300kpc,th}$ for each cluster. According to the notation of Secs.~\ref{sec_regr} and \ref{sec_samp}, the threshold for the $j$-th cluster is
 \beq
 y_{\text{th},j}=l_{\text{th},j}.
 \eeq
The uncertainty $\delta_{y_{\text{th},j}}$ related to the extrapolation procedure is estimated by considering a range of radial profiles with scatter in slope of $\sigma_\beta\sim0.1$, scatter in core radius of $\sigma(r_\text{c}/r_{500})\sim 0.1$, and correlation $\rho_{\beta r_c}\sim 0.66$ as representative of the sample of the 45 bright nearby galaxy clusters in \citet{moh+al99}. The median $\delta_{y_{\text{th}}}$ is $\sim3\%$ but the distribution of values shows a long tail at larger values, so that the mean is $\sim6\%$ and the standard deviation is $\sim7\%$.

\section{Theoretical predictions}
\label{sec_theo}
The self-similar scenario of cluster formation and evolution was first proposed by \citet{kai86} and later on extended and integrated \citep[see e.g.][]{gio+al13}. If gravity is the driving force of structure formation, X-ray quantities follow power-laws. 

The relations for quantities within a fixed physical length differ from the canonical ones within the over-density radius. Here, the reference radius $R$ is constant and it does not scale with the mass. On the other hand, the density within $R$ is not constant and it changes with the mass. Since the density is not a constant multiple of the critical one, the time dependence factor $E_z^2$ connected to the critical density does not enter the relations.

Under the assumptions that clusters are closed boxes and baryons track the total mass,
\beq
M_\text{g} \propto M .  \label{eq_scal_5}
\eeq
If the cluster is near hydrostatic equilibrium, then
\beq
M \sim T\ R \propto T. \label{eq_scal_6}
\eeq
Finally, the X-ray emission in the soft band scales as 
\beq
\epsilon^\text{XXL} \sim \rho_\text{gas}^2,  \label{eq_scal_7}
\eeq
with no appreciable temperature dependence \citep{ett15}. By definition, luminosity can be written as
\beq
L^\text{XXL} \sim \epsilon^\text{XXL} R^3 \propto M_\text{gas}^2.  \label{eq_scal_8}
\eeq
By rearranging Eqs.~(\ref{eq_scal_5}--\ref{eq_scal_8}), we obtain the self-similar scaling relations for properties measured within fixed radii,
\begin{align}
L^\text{XXL} & \propto T^2, \label{eq_scal_1}\\ 
M_\text{g} & \propto T,  \label{eq_scal_2} \\
L^\text{XXL} & \propto M_\text{g}^2.  \label{eq_scal_3}
\end{align}
Reported slopes involving the luminosity are appropriate for the soft band \citep{ett15}. 

Baryonic processes can disrupt the self-similar relations. AGN (Active Galactic Nucleus) feedback or radiative cooling can remove cold, dense gas from the inner regions of low mass clusters, which makes the gas mass vs total mass and luminosity vs total mass relations steeper and the temperature-mass relation shallower \citep{tru+al18}.


Scaling relations can be scattered by a number of processes acting in different directions. Non-thermal sources of gas pressure, temperature inhomogeneity, substructures and clumps, unvirialized bulk motions, and subsonic turbulence play a role \citep{bat+al12,ras+al12}. 

Triaxiality is an additional source of scatter \citep{lim+al13,ser+al13}. Observed signals depend on the orientation of the cluster \citep{gav05,ogu+al05,ser07,se+um11,lim+al13,ser+al13}. For systems whose major axis points toward the observer, which are typically over-represented in signal-limited samples, X-ray luminosities and gas mass derived under the standard assumption of spherical symmetry are over-estimated. On the other hand, the majority of randomly oriented clusters are elongated in the plane of the sky and properties can be under-estimated. 

A certain degree of correlation between intrinsic scatters is in place and has to be considered in multi-property galaxy cluster statistics to properly model the scaling relations \citep{evr+al14,roz+al14c,mau14,man+al15}. Correlations can come from internal structure, formation history, orientation, environment, and uncorrelated structure \citep{ang+al12}.

The X-ray luminosity depends on the assembling history of the clusters \citep{man+al16}. Massive mergers impact the luminosity-mass relation \citep{tor+al04}. The dynamical state of the cluster can cause a positive correlation between luminosity and temperature \citep{man+al16_wtg}. Apart from transient shocks, luminosity and temperature can be depressed in merging clusters where energy in bulk motions has not yet virialized. On the other hand, dynamically relaxed, hot clusters show bright cores with higher than average luminosities and approximately average temperatures. In fact, perturbed or relaxed clusters move coherently along the luminosity-temperature relation \citep{row+al04,har+al08}.

Intra cluster-medium (ICM) processes impact correlations too. Radiative cooling reduce the amount of gas in the smallest systems, and at the same time their total luminosity  \citep{tru+al18}.

\citet{ett15} showed how the normalizations of the scaling relations between the hydrostatic mass and the gas mass, the gas temperature,  the X-ray bolometric luminosity, and the integrated Compton parameter depend upon the gas density clumpiness, the gas mass fraction, and the logarithmic slope of the thermal pressure profile. Scatter in the thermal pressure profile can cause positive correlation between the spectroscopic estimate of the temperature and the X-ray luminosity. Clumpiness induces positive correlation between the luminosity and the gas mass estimated under the hypothesis of a smooth distribution. \citet{ett15} also argued that deviations of the observed slopes from the self-similar expectations can be explained with a mass dependence of the gas mass fraction and of the logarithmic slope of the thermal pressure profile.

X-ray quantities in numerical simulations are positively correlated \citep{sta+al10,tru+al18}. \citet{sta+al10} performed a numerical study of the intrinsic covariance of cluster observables using the Millennium Gas Simulations. They adopted two different physical treatments: shock heating driven by gravity only (GO), or cooling and preheating (PH). The results in \citet{sta+al10} depend on the adopted scheme. They found correlation factors at redshift zero between bolometric luminosity, spectroscopic-like temperature, and gas mass within $r_{500}$ of $\rho_{lt|m}=0.67$ (0.73), $\rho_{lm_\text{g}|m}=0.60$ (0.76), $\rho_{m_\text{g}t|m}=0.42$ (0.37) for the GO (PH) simulation.

\section{Results}
\label{sec_resu}

\begin{table*}
\caption{Observed scaling relations. Conventions and units are as in Section~\ref{sec_regr}. Cols.~1-2: variables of the regression procedure. Cols.~3, 4, and 5: intercept, slope, and time evolution of the scaling relation. Cols.~6-7: scatter of $X$ and its time-evolution. Cols.~8-9: scatter of $Y$ and its time-evolution. Cols.~10-11: correlation between the scatters and its time evolution. If the evolution is not considered ($\gamma_{\sigma_{X|Z}}=\gamma_{\sigma_{Y|Z}}=\gamma_{\rho_{XY|Z}}=0$), scatters and correlations are meant as weighted averages over the redshift range. If the evolution is free, scatters and correlations are intended as local values at $z_\text{ref}=0.01$. Col.~12: self-similar slope ($\beta_\text{ss}$) for observable measured within fixed radii. Values in square brackets correspond to parameters kept fixed in the regression. We report the medians of the marginalized posterior distributions and the 68.3\% probability range. Quoted values are biweight estimators of the marginalized distributions.
}
\label{tab_scaling}
\centering
\resizebox{\hsize}{!} {
\begin{tabular}[c]{l l r@{$\,\pm\,$}l  r@{$\,\pm\,$}l  r@{$\,\pm\,$}l  r@{$\,\pm\,$}l  r@{$\,\pm\,$}l  r@{$\,\pm\,$}l r@{$\,\pm\,$}l  r@{$\,\pm\,$}l  r@{$\,\pm\,$}l l}
\hline
\noalign{\smallskip} 
	\multicolumn{2}{c}{} & \multicolumn{6}{c}{Scaling}& \multicolumn{4}{c}{Intrinsic scatter in $X$}& \multicolumn{4}{c}{Intrinsic scatter in $Y$}& \multicolumn{4}{c}{Scatter correlation} &         \\ 
\noalign{\smallskip} 
	\multicolumn{2}{c}{} & \multicolumn{2}{c}{Intercept}& \multicolumn{2}{c}{Slope}& \multicolumn{2}{c}{Evolution}&\multicolumn{2}{c}{Local scatter}&\multicolumn{2}{c}{Evolution}& \multicolumn{2}{c}{Local scatter}&\multicolumn{2}{c}{Evolution}& \multicolumn{2}{c}{Local correlation} &\multicolumn{2}{c}{Evolution}& 
        \\ 
	\noalign{\smallskip}  
	 $x$ &  $y$& 	\multicolumn{2}{c}{$\alpha_{Y|Z}$}	&	\multicolumn{2}{c}{$\beta_{Y|Z}$}&	\multicolumn{2}{c}{$\gamma_{Y|Z}$}&	\multicolumn{2}{c}{$\sigma_{X|Z,0}$}&	\multicolumn{2}{c}{$\gamma_{\sigma_{X|Z}}$}&	\multicolumn{2}{c}{$\sigma_{Y|Z,0}$}&	\multicolumn{2}{c}{$\gamma_{\sigma_{Y|Z}}$}  &	\multicolumn{2}{c}{$\rho_{XY|Z,0}$}&	\multicolumn{2}{c}{$\gamma_{\rho_{XY|Z}}$}   & 
	 $\beta_\mathrm{ss}$\\
	\hline
	\noalign{\smallskip}  
	$t$	&		$l$	 & 	
	$-$2.15&0.18&   2.78&0.42&    $-$0.48&1.25&
	0.03&0.04&    \multicolumn{2}{c}{[0]}&    0.29&0.08&      \multicolumn{2}{c}{[0]}&  0.20&0.48&   \multicolumn{2}{c}{[0]}&	 2\\
	$t$	&		$l$	 & 	
	$-$2.13&0.18&   2.70&0.42&    $-$0.50&1.29&       
	0.03&0.03&    0.39&0.34&    0.29&0.08&      0.28&0.28&  0.28&0.48&   0.32&0.66&	 2\\
	$t$	&		$m_\text{g}$	 &	
	$-$2.29&0.06&   1.53&0.20&    0.53&0.53&  0.10&0.02&     \multicolumn{2}{c}{[0]}&       0.03&0.03&    \multicolumn{2}{c}{[0]}&  0.64&0.48& 	\multicolumn{2}{c}{[0]}& 1\\
	$t$	&		$m_\text{g}$	 & 	
	$-$2.27&0.05&   1.47&0.17&    0.67&0.52&  0.09&0.02&     0.23&0.25&       0.02&0.02&   0.45&0.34&  0.46&0.52& 	$-$0.05&0.63& 1\\
	$m_\text{g}$      &              $l$	 & 	
	2.32&0.18&   1.91&0.09&    $-$0.44&0.49&    
	0.04&0.02&     \multicolumn{2}{c}{[0]}&    0.12&0.05&     \multicolumn{2}{c}{[0]}& 0.89&0.20& 	 \multicolumn{2}{c}{[0]}& 2 \\
	$m_\text{g}$      &              $l$	 & 	
	2.43&0.14&   1.97&0.07&    $-$0.73&0.42&    
	0.02&0.01&    0.42&0.33&    0.06&0.03&     0.42&0.33& 0.31&0.58& 	 0.04&0.63&		2\\
	\hline
	\end{tabular}
	}
\end{table*}

\begin{table}[]
\caption{Scaling relations and intrinsic scatters from the multi-response regression as described in Section~\ref{sec_mult}. The gas mass $m_\text{g}$, the temperature $t$, and the luminosity $l$ act as $y_1$, $y_2$, and $y_3$, respectively. $Y_1$ is an unbiased but scattered proxy of $Z$. Scatters and correlations are meant as weighted averages over the redshift range. Values in square brackets are kept fixed in the regression. The last column report the self-similar expectations. Quoted values are biweight estimators of the marginalized distributions.
}
\label{tab_scaling_multi}
\centering
\resizebox{\hsize}{!} {
\begin{tabular}[c]{l l l  r@{$\,\pm\,$}l c}
        \noalign{\smallskip}  
        \hline
         \noalign{\smallskip}  
         Description & Relation & Parameter & \multicolumn{2}{c}{Results} & S-S\\
	\hline
	\noalign{\smallskip}  
	intercept &		$M_\text{gas,300kpc}$-$M_\text{gas,300kpc}$& $\alpha_{m_\text{g}^{(Y)}|m_\text{g}^{(Z)}}$&	\multicolumn{2}{c}{[0]}	&	--\\
	intercept &		$T_\text{300kpc}$-$M_\text{gas,300kpc}$& $\alpha_{t^{(Y)}|m_\text{g}^{(Z)}}$& 				1.60&0.17	&	--\\
	intercept &		$L^\text{XXL}_\text{300kpc}$-$M_\text{gas,300kpc}$& $\alpha_{l^{(Y)}|m_\text{g}^{(Z)}}$& 		2.39&0.13	&	--\\
	slope&			$M_\text{gas,300kpc}$-$M_\text{gas,300kpc}$& $\beta_{m_\text{g}^{(Y)}|m_\text{g}^{(Z)}}$&	 \multicolumn{2}{c}{[1]}	&	1\\
	slope&			$T_\text{300kpc}$-$M_\text{gas,300kpc}$& $\beta_{t^{(Y)}|m_\text{g}^{(Z)}}$&  				0.71&0.09	&	1\\
	slope&			$L^\text{XXL}_\text{300kpc}$-$M_\text{gas,300kpc}$& $\beta_{l^{(Y)}|m_\text{g}^{(Z)}}$& 		1.94&0.07	&	2\\
	time-evolution&		$M_\text{gas,300kpc}$-$M_\text{gas,300kpc}$&  $\gamma_{m_\text{g}^{(Y)}|m_\text{g}^{(Z)}}$&  \multicolumn{2}{c}{[0]}	&	0\\
	time-evolution&		$T_\text{300kpc}$-$M_\text{gas,300kpc}$& $\gamma_{t^{(Y)}|m_\text{g}^{(Z)}}$&  			0.68&0.09	&	0\\
	time-evolution&		$L^\text{XXL}_\text{300kpc}$-$M_\text{gas,300kpc}$& $\gamma_{l^{(Y)}|m_\text{g}^{(Z)}}$& 	0.15&0.55	&	0\\
	intrinsic scatter&	$M_\text{gas,300kpc}$|$Z$& $\sigma_{m_\text{g}^{(Y)}|Z}$&  							0.03&0.01	&	--\\
	intrinsic scatter&	$T_\text{300kpc}$|$Z$ & $\sigma_{t^{(Y)}|Z}$&										0.09&0.02	&	--\\
	intrinsic scatter&	$L^\text{XXL}_\text{300kpc}$|$Z$  & $\sigma_{l^{(Y)}|Z}$& 								0.05&0.03	&	--\\
	scatter correlation& $T_\text{300kpc}$-$M_\text{gas,300kpc}$| $Z$& $\rho_{m_\text{g}^{(Y)}t^{(Y)}|Z}$&  			0.35&0.52	&	--\\
	scatter correlation& $L^\text{XXL}_\text{300kpc}$-$M_\text{gas,300kpc}$| $Z$& $\rho_{m_\text{g}^{(Y)}l^{(Y)}|Z}$& 	0.40&0.43	&	--\\
	scatter correlation& $L^\text{XXL}_\text{300kpc}$-$T_\text{300kpc}$| $Z$& $\rho_{t^{(Y)}l^{(Y)}|Z}$&				0.07&0.70	&	--\\
	\hline
	\end{tabular}
	}
\end{table}

As reference analysis, we consider the statistical approach detailed in Section~\ref{sec_regr} in which X-ray observables are analyzed in pairs. Since the procedure is symmetric and both $X$ and $Y$ are affected by scatter, the results do not change whether we associate one observable to either $X$ or $Y$. The analysis was performed with the \textsc{R}-package \textsc{LIRA}\footnote{The package  \textsc{LIRA} (LInear Regression in Astronomy) is publicly available from CRAN (Comprehensive R Archive Network) at \url{https://cran.r-project.org/web/packages/lira/index.html}.}. As reference redshift, we consider $z_\text{ref}=0.01$. Results for the scaling parameters and the intrinsic scatters are summarized in Table~\ref{tab_scaling}. 

As an alternative method, we fitted the three X-ray observables at once, see Section~\ref{sec_mult}. The Bayesian hierarchical multi-response model was sampled with JAGS\footnote{JAGS (Just Another Gibbs Sampler) is a program for Bayesian data analysis publicly available at \url{http://mcmc-jags.sourceforge.net/}.}. 

A critical aspect in Bayesian data analysis is the computational efficiency of the sampling of the posterior probability distribution. Gibbs sampling or other methods can efficiently constrain the distribution \citep{kel07,man16}, but problems may arise for non standard distributions. This is the case of the truncated multinormal distributions presented in Sec.~\ref{sec_mult}. To circumvent the problem, we neglected the truncation in Eq.~(\ref{eq_bug_multi_3}) but we still considered the truncation in Eq.~(\ref{eq_bug_multi_2}), which is much easier to sample since the uncertainty covariance matrix is known and fixed. We tested that this approximate scheme can still correct for the main effects of the Malmquist bias. 

In our multi-response analysis, the gas mass acts as the $Y_1$ variable. As for the analysis in pairs, the choice of the $Y_1$ variable does not affect the measurement of the scatters. Results are summarized in Table~\ref{tab_scaling_multi}. The agreement between the two alternative regression methods is substantial. Values listed in Tables~\ref{tab_scaling} and \ref{tab_scaling_multi} are in base 10 logarithm.

\subsection{Scaling}
\label{sec_scal}

\begin{figure}
\begin{tabular}{c}
\resizebox{\hsize}{!}{\includegraphics{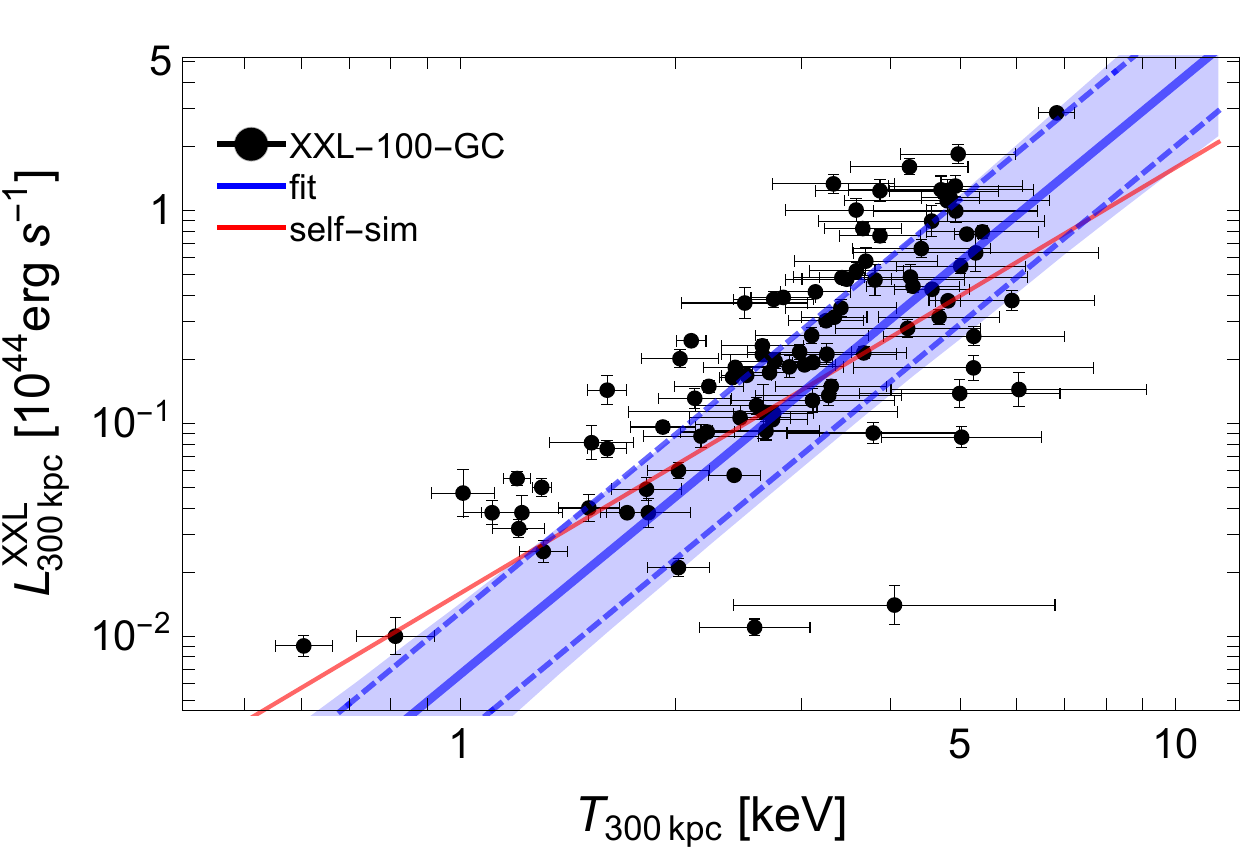}} \\
\resizebox{\hsize}{!}{\includegraphics{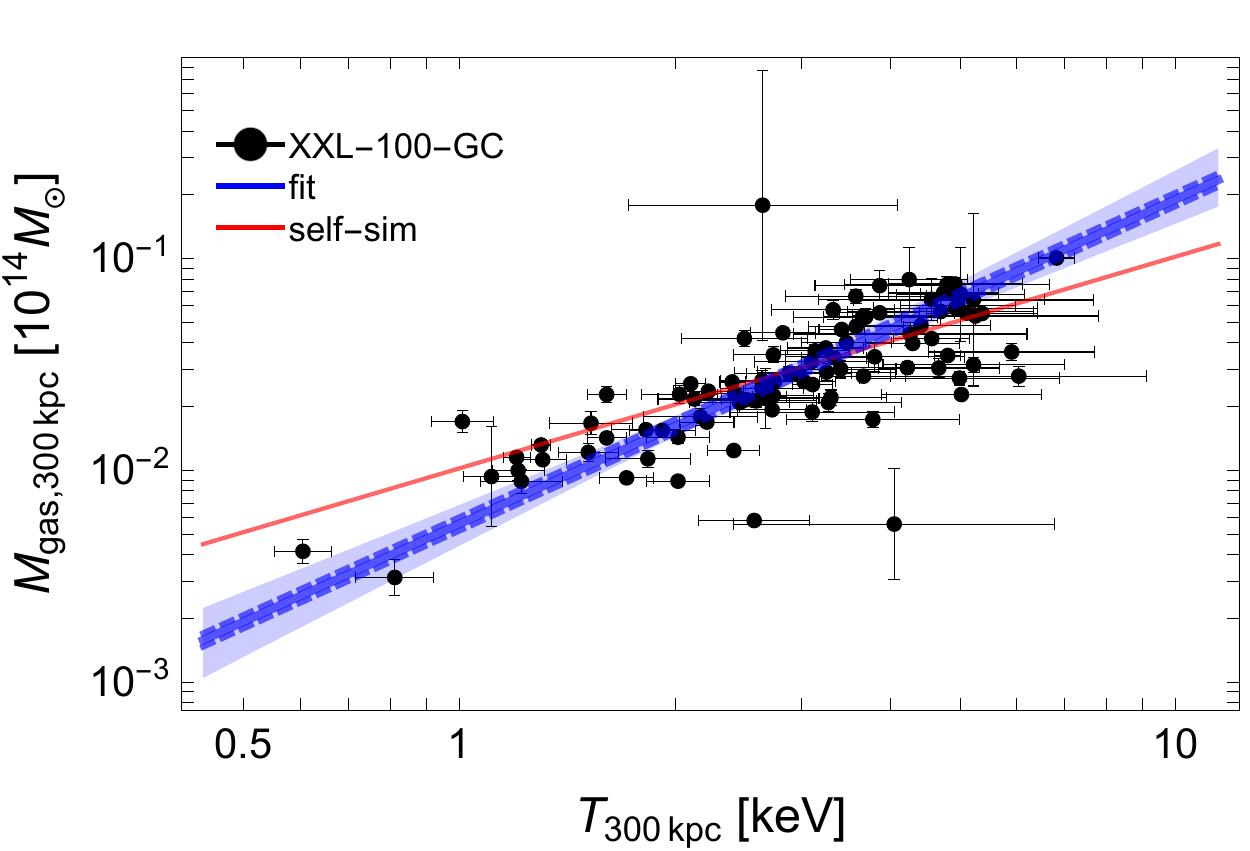}} \\ 
\resizebox{\hsize}{!}{\includegraphics{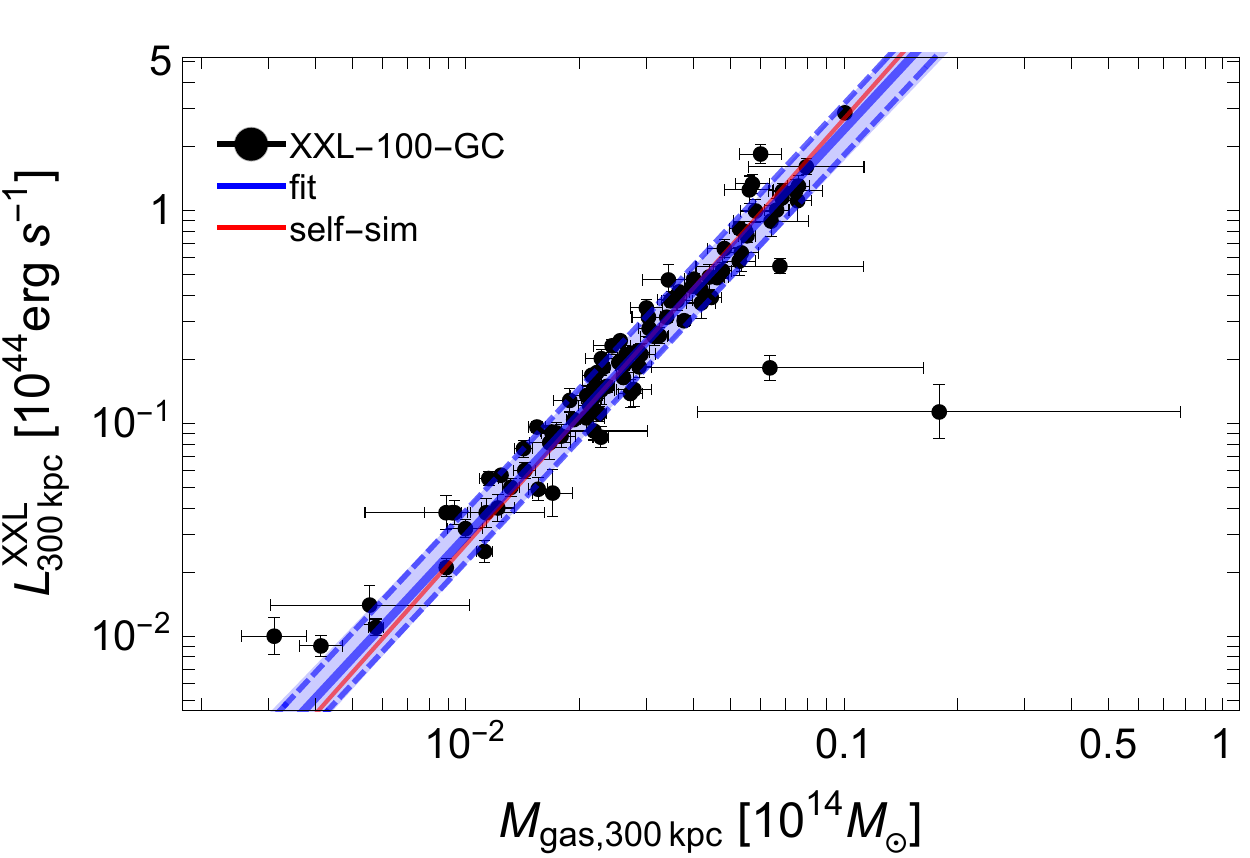}} \\
\end{tabular}
\caption{Scaling relations. The black points mark the data, the blue lines represent the fitted scaling relation at the median redshift. The dashed blue lines show the median scaling relation (full blue line) plus or minus the intrinsic scatter $\sigma_{Y|Z}$. The shaded blue region encloses the $68.3$ per cent confidence region around the median due to uncertainties on the scaling parameters. The red line shows the self-similar prediction. {\it Top panel}: scaling between luminosity and temperature, $l$-$t$. {\it Middle panel}: scaling between gas mass and temperature, $m_\text{g}$-$t$. {\it Bottom panel}: scaling between luminosity and gas mass, $l$-$m_\text{g}$.}
\label{fig_scaling}
\end{figure}

Results of the reference regression analysis are reported in Table~\ref{tab_scaling}. Measured gas mass, temperature, and luminosity are aligned quite well, see Fig.~\ref{fig_scaling}. The marginalized 2D posteriori probabilities are showed in App.~\ref{app_2D}.

The $l$-$t$ and $m_\text{g}$-$t$ relations are steeper than the self-similar predictions. This is confirmed by the multi-response analysis, see Table~\ref{tab_scaling_multi}. There is no statistical evidence for time evolution, i.e. the $\gamma$ parameters are consistent with zero. 

Parameter degeneracies can be analyzed with the 2D probability functions. As well known, slope and intercept are anti-correlated (see top-left panels in Figs.~\ref{fig_lx_tx_PDF_2D}, \ref{fig_mg_tx_PDF_2D}, and \ref{fig_lx_mg_PDF_2D}).

Measured slopes are consistent with a prominent role of radiative cooling and AGN feedback in low mass systems, see Sec.~\ref{sec_theo}. Stellar formation consumes the reservoirs of cold gas in the inner regions. Even though AGN feedback balances against over-cooling, it expels gas from the cluster core, which also reduces the luminosity and the total gas supply. As a result of these baryonic processes, which are most effective in low mass systems, the $l$-$t$ and $m_\text{g}$-$t$ relations are steeper than the self-similar predictions, in agreement with our results.


\subsection{Intrinsic scatters}

\begin{figure}
\begin{tabular}{c}
\resizebox{\hsize}{!}{\includegraphics{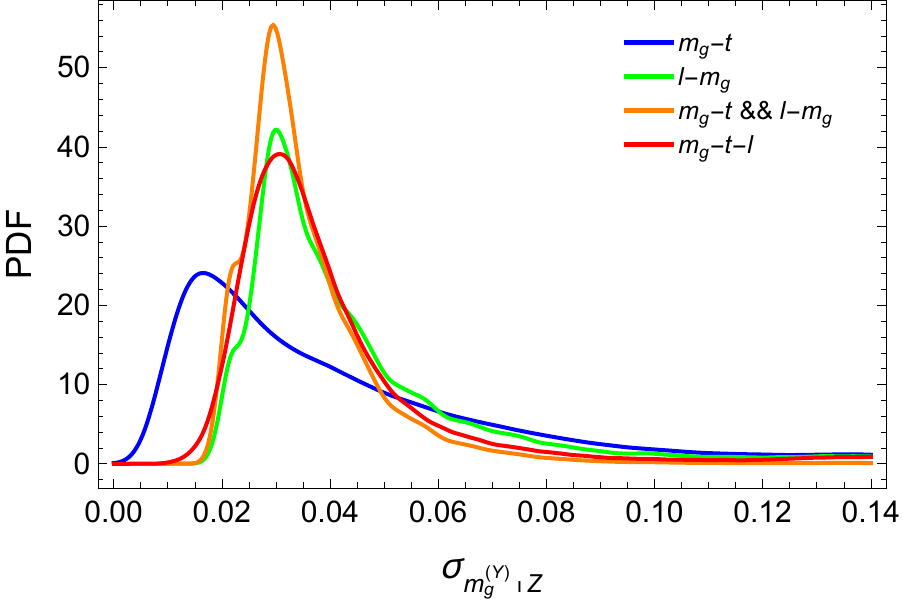}} \\
\resizebox{\hsize}{!}{\includegraphics{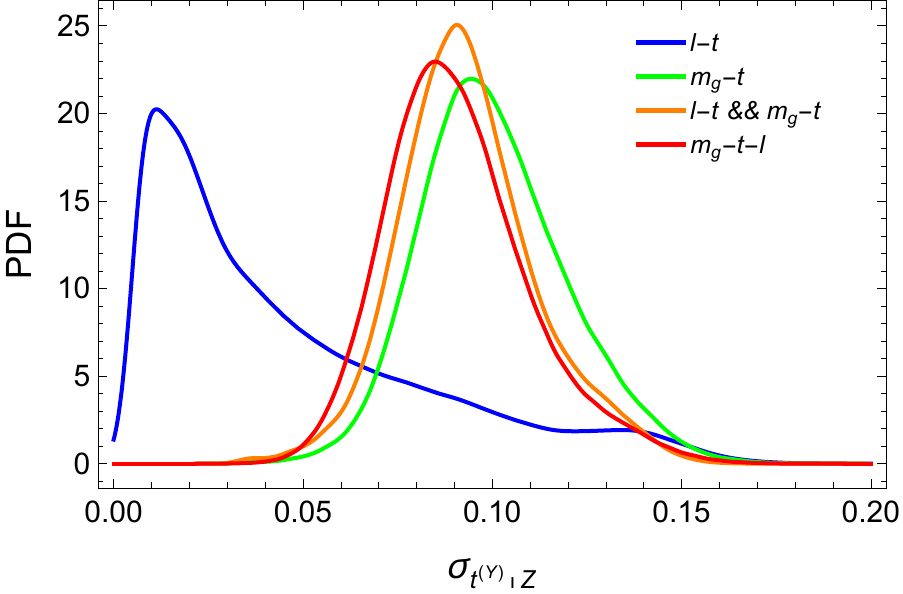}} \\ 
\resizebox{\hsize}{!}{\includegraphics{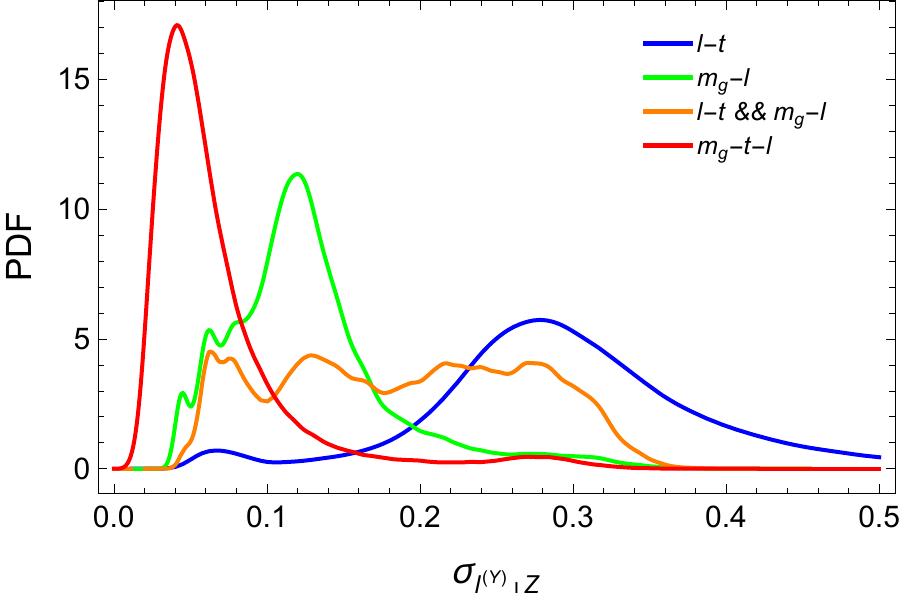}} \\
\end{tabular}
\caption{Inferred probability density functions of the conditional intrinsic scatters with respect to the mass substitute. The blue and the green lines show the results based on the fits of pairs; the orange and the red lines show the densities obtained with the joint analysis of pair fits or with the multi-response regression, respectively. {\it Top panel}:  gas mass intrinsic scatter. {\it Middle panel}: temperature intrinsic scatter. {\it Bottom panel}: luminosity intrinsic scatter.}
\label{fig_pdf_sigma}
\end{figure}

The gas mass is the less scattered proxy, $\langle \sigma_{m_\text{g}^{(Y)}|Z}\rangle=0.04\pm0.01$ ($8.2\pm3.0\%$), followed by temperature, $\langle\sigma_{t^{(Y)}|Z}\rangle=0.09\pm0.02$ ($21.5\pm4.2\%$) and luminosity, $\langle\sigma_{l^{(Y)}|Z}\rangle=0.19\pm0.08$ ($43.6\pm18.5\%$), see Fig.~\ref{fig_pdf_sigma}. In the reference analysis, the three X-ray quantities were analyzed in pairs, and we have two measurements of each scatter. The scatter values quoted above were obtained as the weighted mean and the standard deviation of the probability density obtained by combining the results from the two regressions summarized in Table~\ref{tab_scaling}. 

The reference results (plotted as orange lines in Fig.~\ref{fig_pdf_sigma}) are fully consistent with the results from the multi-response analysis (plotted as red lines in Fig.~\ref{fig_pdf_sigma}), when we obtained $\langle\sigma_{m_\text{g}^{(Y)}|Z}\rangle=0.04\pm0.02~(9.4\pm5.2\%)$, $\langle\sigma_{t^{(Y)}|Z}\rangle=0.09\pm0.02~(20.9\pm4.4\%)$ and $\langle\sigma_{l^{(Y)}|Z}\rangle=0.07\pm0.05~(15.6\pm11.7\%)$. Due to skewness, the mean and standard deviation slightly differ from the biweight estimators quoted in Table~\ref{tab_scaling_multi}.

The probability distribution of the luminosity intrinsic scatter extends over a significantly larger range than gas mass and temperature with a tail at the upper end, see Fig.~\ref{fig_pdf_sigma}. The standard deviation, i.e. the quoted uncertainty, is then larger than for $m_\text{g}$ and $t$. 

Furthermore, the analysis of luminosity is more directly affected by Malmquist bias, which we treat as a threshold in the observed luminosity, see Sec.~\ref{sec_cova}. Thresholds are expressed in terms of probability distributions, see Eqs.~(\ref{eq_mb_1},~\ref{eq_mb_2},~\ref{eq_bug_multi_4},~\ref{eq_bug_multi_5}), which affects the precision within which regression parameters are recovered.

The intrinsic scatter is best constrained when at least one low scatter proxy, e.g. the gas mass, is included in the fitting procedure. As seen from the comparison with the multivariate analysis in the middle and lower panels of Fig.~\ref{fig_pdf_sigma}, the $l$-$t$ fitting overestimates the larger scatter, i.e. $\sigma_{l^{(Y)}|Z}$, and underestimates the smaller one, i.e. $\sigma_{t^{(Y)}|Z}$. These two scatters are in fact anti-correlated, see Fig.~\ref{fig_lx_tx_PDF_2D}. The posterior probability distributions are however compatible.

The temperature is intrinsically less scattered than the luminosity but is measured with a larger uncertainty. The two effects partially counter-balance and make the two proxies nearly as effective.

The time evolution of scatters is not well constrained. Parameter values are affected by large statistical uncertainties and are consistent with zero.

Our estimates of the scatter can be slightly overestimated since we compare deprojected quantities measured within the sphere, e.g. the gas mass, to projected quantities measured within the cylinder, e.g. the temperature and the luminosity. The dispersion associated to the non-universality of the density profiles slightly inflates the measured scatters.

In our analysis, we assume that the variable $X$ is an unbiased proxy, i.e. Eq.~(\ref{eq_bug_1}) reduces to $X_Z = Z$. The role of the $X$ was covered by either the temperature or the gas mass. The gas mass also worked as $Y_1$, i.e. the analog of $X$ in the multi-response analysis. These assumptions are in line with our expectations for quantities measured within a fixed radius, when $M_\text{g} \propto M$ and $T \propto M$, see Sec.~\ref{sec_scal}. However, the determination of the conditional scatters and of the correlation factors is independent of the assumption on $X$. In fact, the estimates of scatters and slope are just weakly correlated, see Figs.~\ref{fig_lx_tx_PDF_2D}, \ref{fig_mg_tx_PDF_2D}, and \ref{fig_lx_mg_PDF_2D} and \citetalias{se+et15_comalit_IV}. We further validate the stability of our results by considering tilted relations between $X$ and $Z$, e.g. $X_Z = 3\times Z$, or interchanging the roles of $Y$ and $X$. Notwithstanding the different values of $\beta_{X|Z}$, the results for scatter covariance matrix are the same.

\subsection{Scatter correlations}

Our analysis suggests that the intrinsic scatters of gas mass, temperature, and luminosity are positively correlated, i.e. clusters of a given mass which are over-luminous have high temperature and an excess of gas mass. The probability distributions are peaked towards $\rho_{XY|Z} \sim 1$. Strong correlations are slightly preferred but the evidence is marginal due to the large statistical uncertainties. In fact, positive correlation is inferred at just the 1-$\sigma$ confidence level. Lower values of the correlation cannot be excluded, see the last rows of Figs.~\ref{fig_lx_tx_PDF_2D}, \ref{fig_mg_tx_PDF_2D}, and \ref{fig_lx_mg_PDF_2D}. This is confirmed by the multi-response analysis, see Fig.~\ref{fig_mg_tx_lx_PDF_2D}.

The intrinsic correlation $\rho_{XY|Z}$ is partially degenerate with the slope of the relation $\beta_{Y|Z}$, see Figs.~\ref{fig_lx_tx_PDF_2D}, \ref{fig_mg_tx_PDF_2D}, and \ref{fig_lx_mg_PDF_2D}.

Positive correlation between the properties of the ICM is expected as a results of formation history and dynamical state, see Sec.~\ref{sec_theo}. Dynamically relaxed clusters are usually hotter and more luminous than merging systems of comparable total mass, where the ICM has yet to virialize  \citep{row+al04,har+al08,man+al16_wtg}. Positive correlation between the luminosity and the gas mass estimated under the hypothesis of a smooth distribution can be induced by clumpiness and the degree of regularity of the system \citep{ett15}. Luminosity and gas mass can be overestimated in triaxial clusters elongated along the line of sight.

Our results support positive correlations even though the large statistical uncertainties ($\delta\rho\sim0.5$, see Tale~\ref{tab_scaling_multi}) prevent to distinguish between extreme scenarios ($\rho~\la 1$) and mild effects of co-evolution ($\rho~\ga 0$).

\subsection{Distribution of the selected clusters}

\begin{figure}
\begin{tabular}{c}
\resizebox{\hsize}{!}{\includegraphics{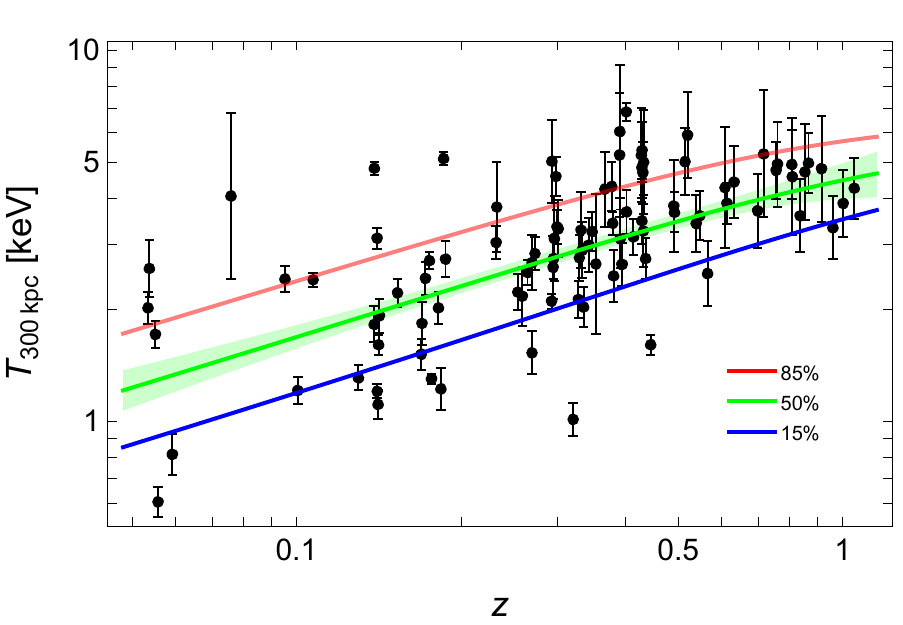}} \\
\resizebox{\hsize}{!}{\includegraphics{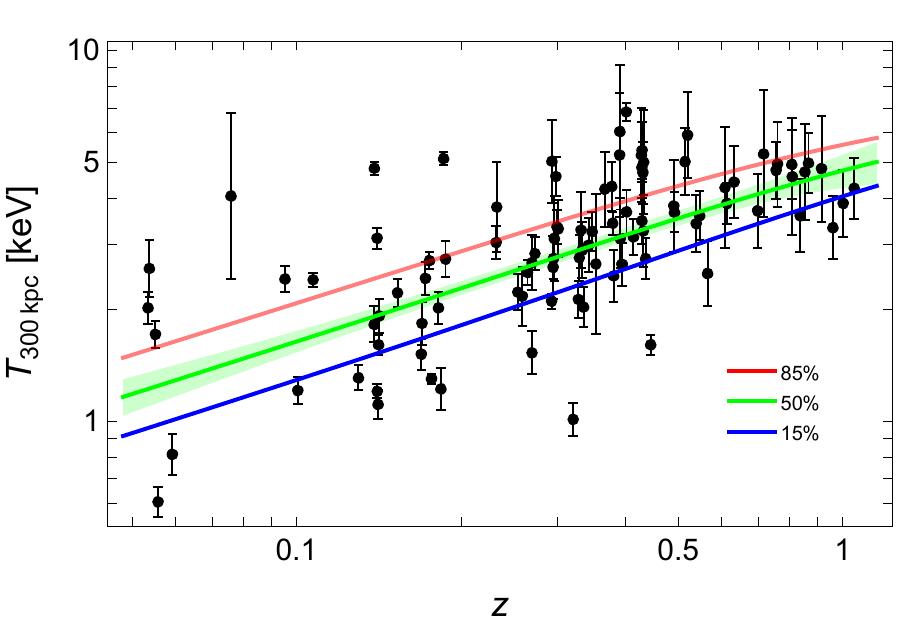}} \\
\resizebox{\hsize}{!}{\includegraphics{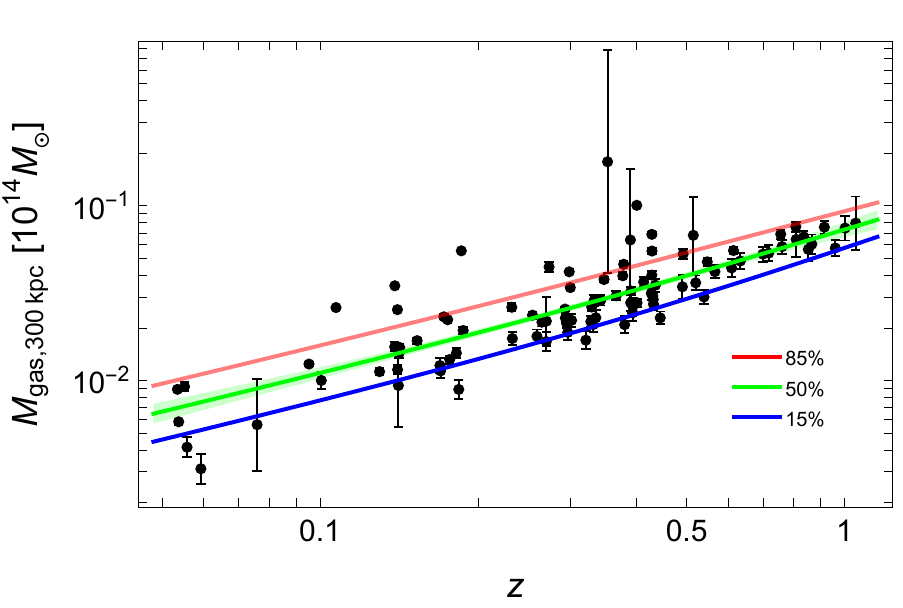}} \\
\end{tabular}
\caption{Time dependence of the X-ray observables of the selected clusters. Measured values are shown as a function of redshift (black points). The full lines plot the value of the unscattered observable below which a given fraction of the selected sample is contained. From top to bottom, the red, green, and blue lines show the 85, 50 ($\mu_Z$), and 15 per cent levels, respectively. The shaded green region encloses the 68.3 per cent confidence region around $\mu_Z(z)$ due to uncertainties on the parameters. {\it Top panel}: time dependence of the temperatures as inferred from the $l$-$t$ relation.  Temperatures are in units of keV. {\it Middle panel}: time dependence of the distribution of temperatures of the selected clusters as inferred from the $m_\text{g}$-$t$ relation. {\it Bottom panel}: time dependence of the distribution of gas masses of the selected clusters as inferred from the $l$-$m_\text{g}$ relation. Masses are in units of $10^{14}M_\odot$.}
\label{fig_Z_evolution}
\end{figure}

\begin{table*}
\caption{Intrinsic distributions of the observed samples modeled as Gaussian functions. Conventions and units are as in Sec.~\ref{sec_regr}. Quoted values are biweight estimators of the marginalized distributions.}
\label{tab_Z_func}
\centering
\begin{tabular}[c]{l  r@{$\,\pm\,$}l  r@{$\,\pm\,$}l  r@{$\,\pm\,$}l  r@{$\,\pm\,$}l  r@{$\,\pm\,$}l }
	\hline
	\noalign{\smallskip}  
	 & \multicolumn{6}{c}{Mean} & \multicolumn{4}{c}{Dispersion} \\ 
	\noalign{\smallskip}  
	Sample &	\multicolumn{2}{c}{$\mu_{Z,0}$}&	\multicolumn{2}{c}{$\gamma_{\mu_Z,D}$}&	\multicolumn{2}{c}{$\gamma_{\mu_Z,F_z}$}&	\multicolumn{2}{c}{$\sigma_{Z,0}$}&	\multicolumn{2}{c}{$\gamma_{\sigma_{Z}}$}  \\
	\noalign{\smallskip}  
	\hline
	\noalign{\smallskip}  
	$l$  -$t$ &			$-$0.22&0.11&    0.44&0.10&    $-$0.43&0.56&    0.15&0.02&   $-$0.72&0.32 \\
	$m_\text{g}$-$t$ &	$-$0.25&0.10&    0.46&0.08&    $-$0.32&0.49&    0.10&0.02&    $-$0.80&0.23 \\
	$l$  -$m_\text{g}$&	$-$2.68&0.12&    0.70&0.10&       0.05&0.54&    0.16&0.02&   $-$0.81&0.22 \\
	\hline
	\end{tabular}
\end{table*}

\begin{figure*}
\begin{tabular}{cc}
\includegraphics[width=8.2cm]{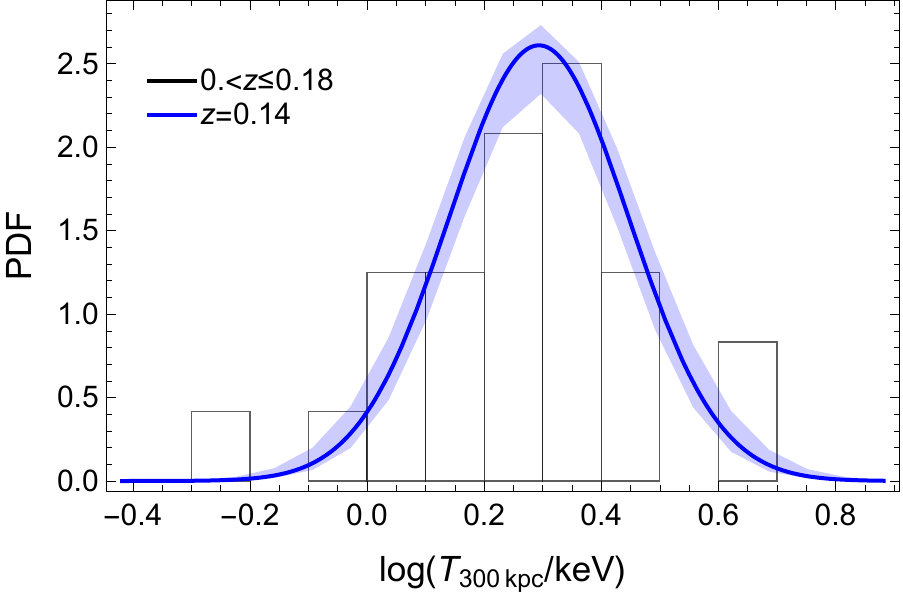} & 
\includegraphics[width=8.2cm]{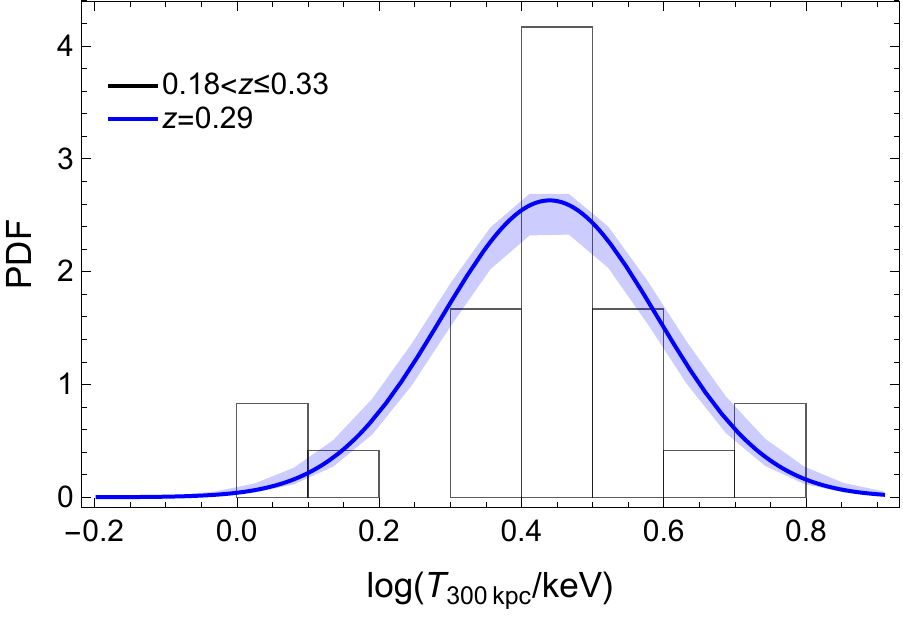} \\
\includegraphics[width=8.2cm]{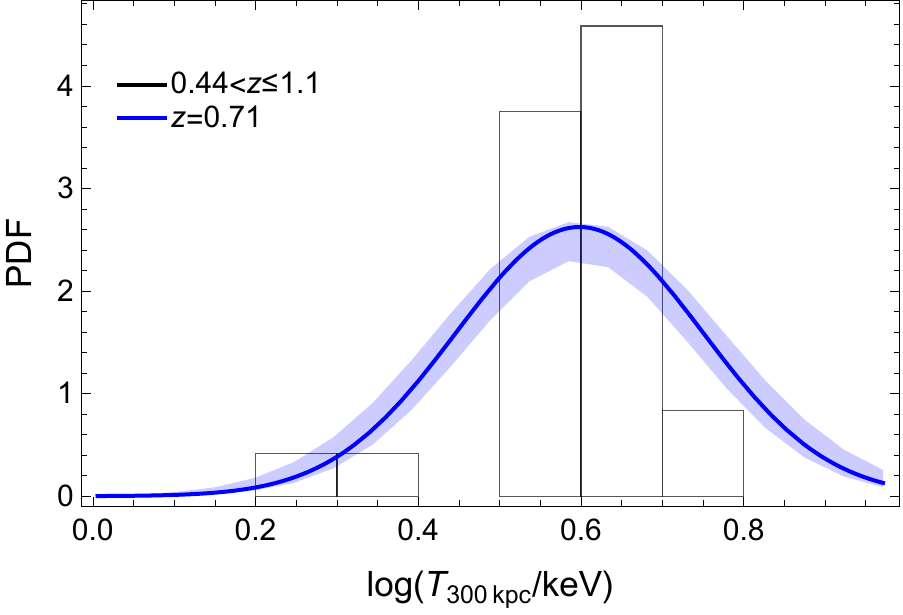} &
\includegraphics[width=8.2cm]{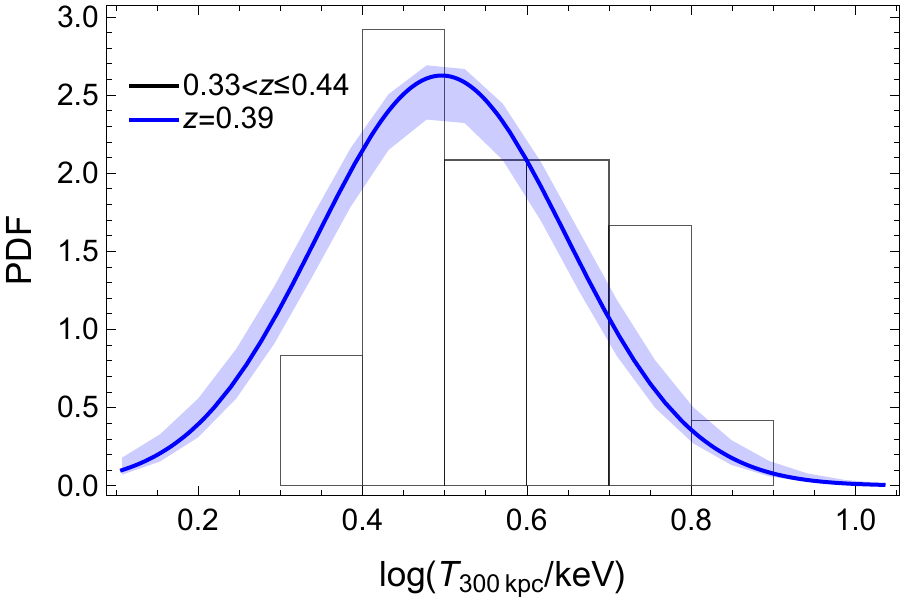} \\
\end{tabular}
\caption{Temperature function of the clusters from the $l$-$t$ analysis in four redshift bins. The black histogram groups the observed temperatures. The blue line is the normal approximation estimated from the regression at the median redshift in the bin. The shaded blue region encloses the 68.3 per cent probability region around the median relation due to uncertainties on the parameters. The function for the observed temperatures is estimated from the regression output, i.e. the function of the unscattered temperatures, by smoothing the prediction with a Gaussian whose variance is given by the quadratic sum of the intrinsic scatter of the (logarithmic) temperature with respect to the unscattered temperature and the median observational uncertainty. Redshift increases clockwise from the top left to the bottom left panel. The median and the boundaries of the redshift bins are indicated in the legends of the respective panels.}
\label{fig_Z_histo_lx_tx}
\end{figure*}

\begin{figure*}
\begin{tabular}{cc}
\includegraphics[width=8.2cm]{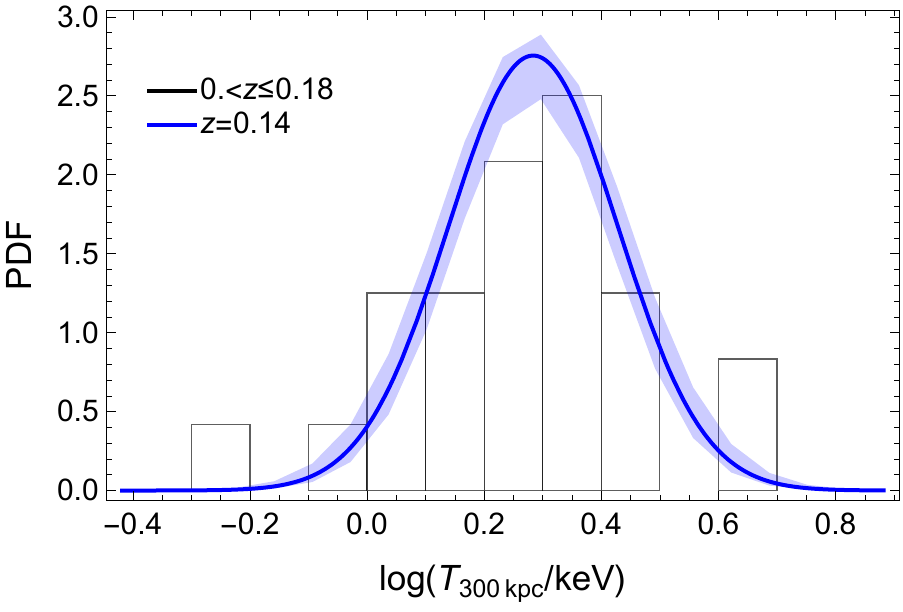} &
\includegraphics[width=8.2cm]{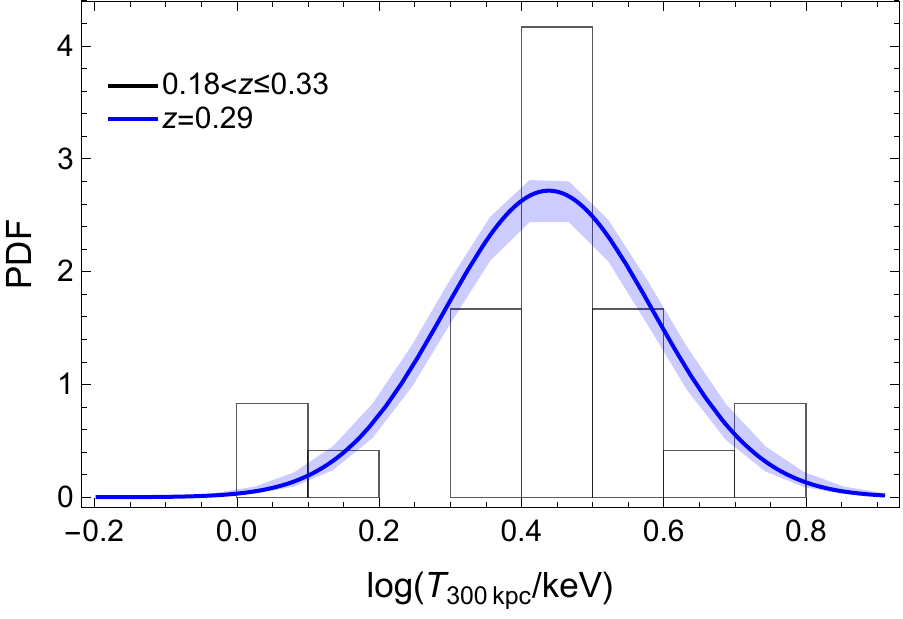} \\
\includegraphics[width=8.2cm]{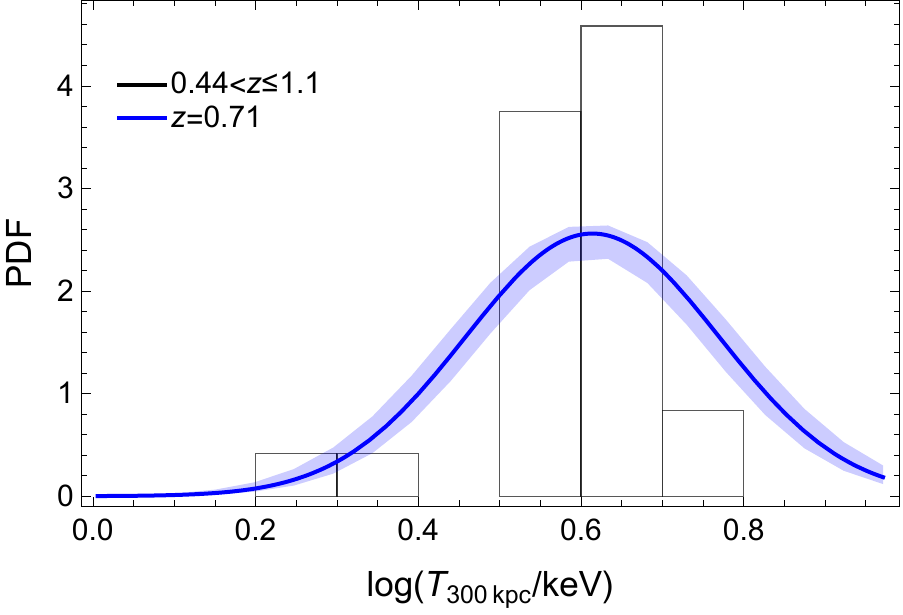} &
\includegraphics[width=8.2cm]{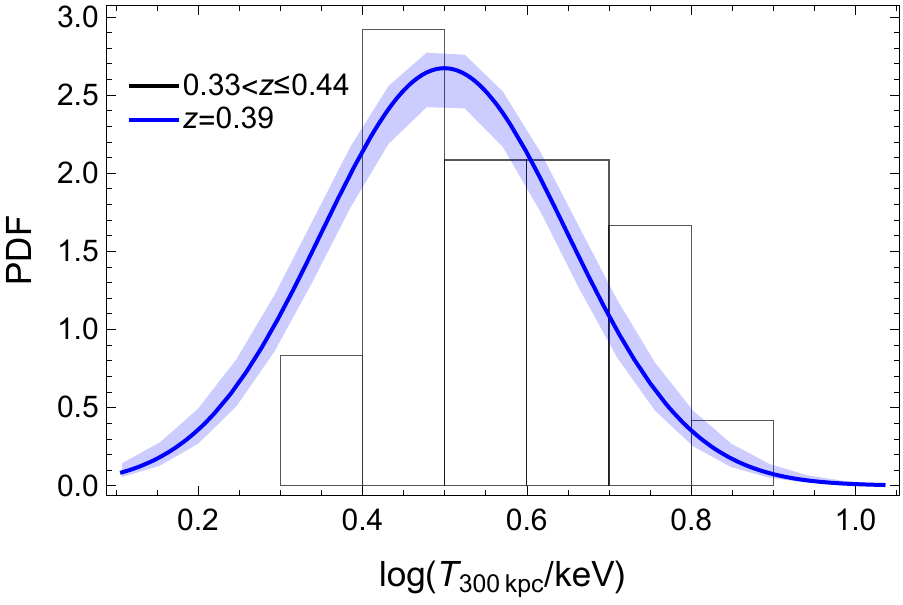} \\
\end{tabular}
\caption{Temperature function of the clusters from the $m_\text{g}$-$t$ analysis in four redshift bins. Lines and conventions are as in Fig.~\ref{fig_Z_histo_lx_tx}.}
\label{fig_Z_histo_mg_tx}
\end{figure*}

\begin{figure*}
\begin{tabular}{cc}
\includegraphics[width=8.2cm]{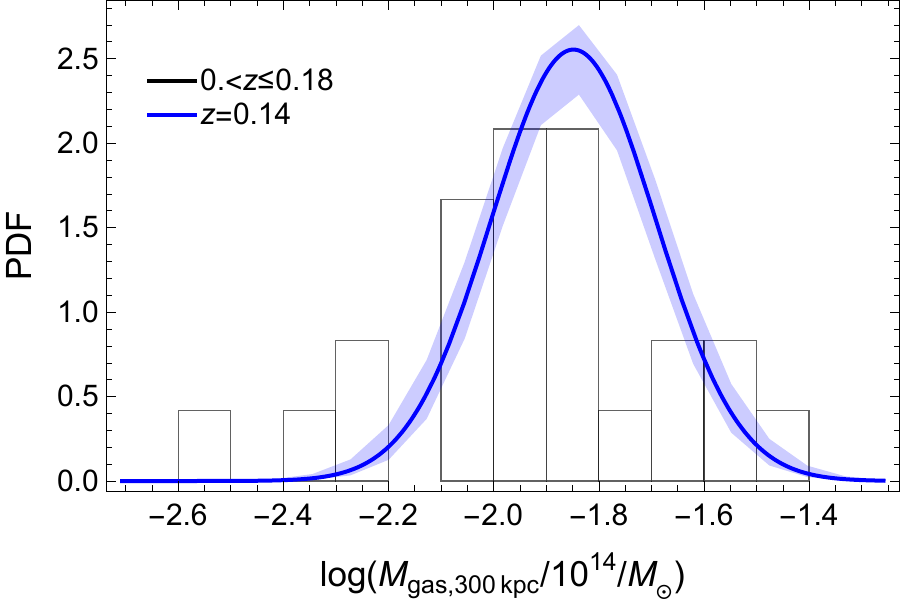} &
\includegraphics[width=8.2cm]{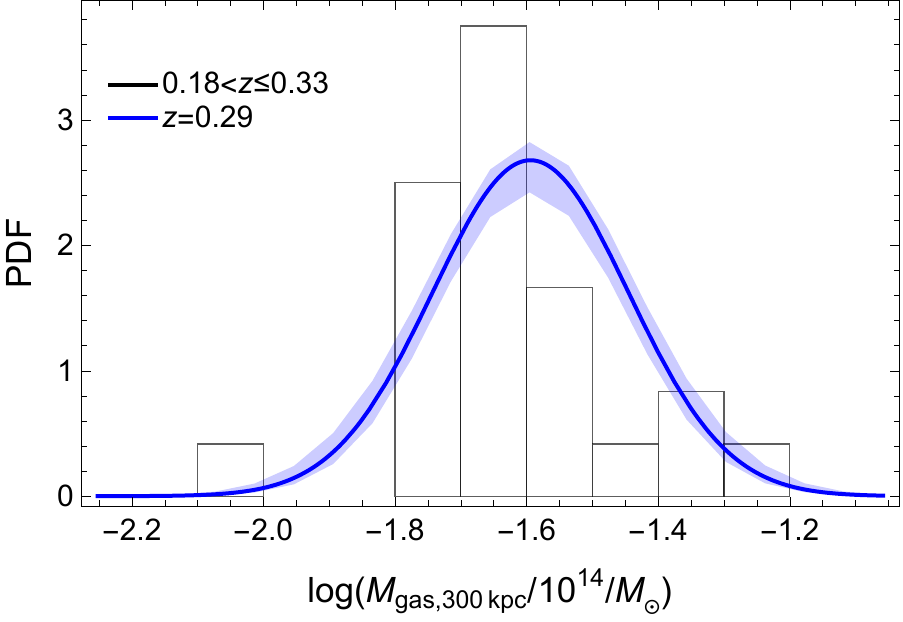} \\
\includegraphics[width=8.2cm]{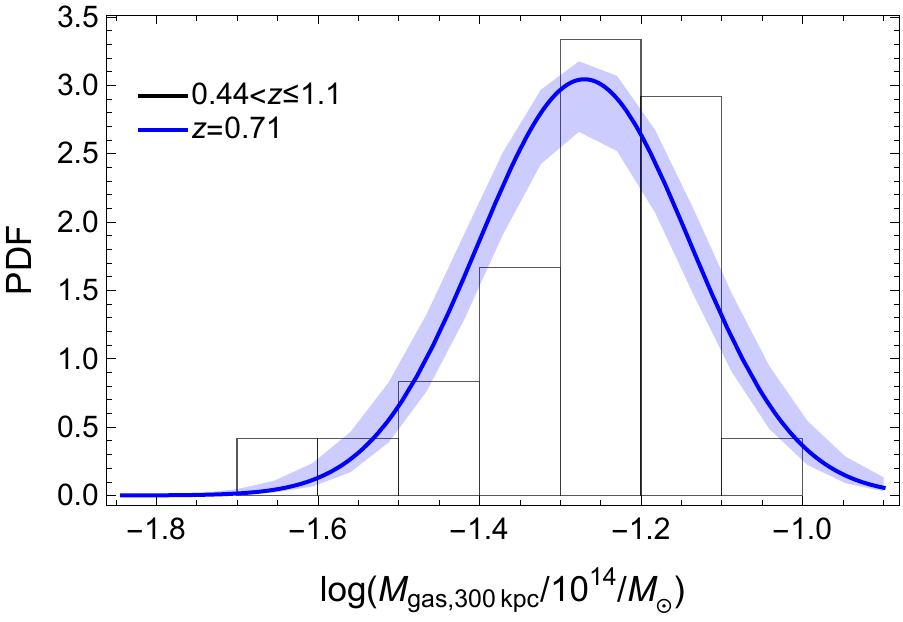} &
\includegraphics[width=8.2cm]{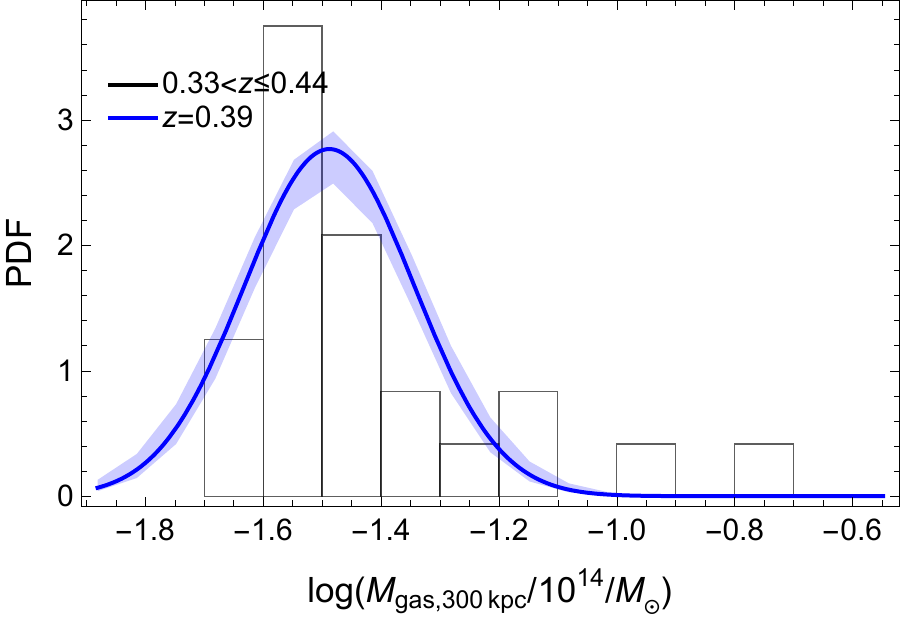} \\
\end{tabular}
\caption{Gas mass function of the clusters from the $l$ -$m_\text{g}$ analysis in four redshift bins. Lines and conventions are as in Fig.~\ref{fig_Z_histo_lx_tx}.}
\label{fig_Z_histo_lx_mg}
\end{figure*}

As expected for flux-limited selected samples, we find that the distributions of gas mass and temperature evolve with redshift, with more massive clusters preferentially included at high redshifts, see Table~\ref{tab_Z_func} and Fig.~\ref{fig_Z_evolution}. These distributions are found as a result of the regression procedure, which does not exploit the knowledge of the mass completeness function obtained by numerical simulations, see Sec.~\ref{sec_sele}, but recovers the distributions from the data.

The Gaussian function provides a good approximation to the redshift evolving distributions, see Figs.~ \ref{fig_Z_histo_lx_tx}, \ref{fig_Z_histo_mg_tx}, and \ref{fig_Z_histo_lx_mg}. The temperature distributions derived from the analysis of the $l$-$t$ and of the $m_\text{g}$-$t$ are fully consistent.

As far as the tails of the covariate distributions are accounted for, conditional scatters and parameters of the scaling relations are well recovered \citep{kel07,ser16_lira}. In fact, a simple normal distribution provides reliable results since it can account at the same time for the selection effects which penalizes low signal clusters and for the rarity of high mass clusters \citep{li+hu05,se+et15_comalit_IV}. A covariate distribution following the halo mass function would fail to reproduce the suppression at the low mass end.

\section{Previous results}
\label{sec_prev}

Intrinsic scatter and correlation can be mass and time-dependent \citep{tru+al18}, and results from different samples should be compared cautiously. We have also to consider differences in measurements and definitions, mostly when comparing relations for either core-excised or core-included quantities. Furthermore, our sample extends to small groups whereas most of the previous analyses considered more massive clusters. Finally, the present level of statistical uncertainties is too large to make firm conclusions on apparent disagreements. However, the positive correlation found with the analysis of XXL-100-GC is consistent with previous results. \citet{mau14} applied the PICACS model to two X-ray samples of clusters with $T \ga2~\text{keV}$ with measured core-excised temperatures, gas masses, and either hydrostatic masses or luminosities. Quantities were measured within $r_{500}$. The analysis suggested a positive correlation between the intrinsic scatter of $T$ and $M_\text{g}$ ($\rho_{m_\text{g}t|m}=0.31\pm0.30$), and between $T$ and the core excluded bolometric luminosity ($\rho_{tl_\text{ce}|m}=0.37\pm0.30$). A strong and significant correlation between the scatter in $M_\text{g}$  and $L_\text{X}$ was found ($\rho_{l_\text{ce}m_\text{g}|m}=0.85\pm0.14$). 

\citet{man+al15} constrained the cosmological parameters through the analysis of the mass function of a sample of X-ray selected massive clusters ($T \ga 4~\text{keV}$) detected in the ROSAT All-Sky Survey. They used follow-up measurements of soft band X-ray luminosity, temperature, and gas mass within $r_{500}$ and weak gravitational lensing measurements of a sub-sample of massive clusters. In the process, they assumed the gas mass scatter as being uncorrelated, i.e. they fixed the off-diagonal covariance terms involving $m_\text{g}$ to zero, and measured the correlation $\rho_{tl|m}=0.11\pm0.19$. However, just using a larger amount of follow-up measurements and an updated calibration for X-ray observations, \citet{man+al16_wtg} found positive strong correlation in the intrinsic scatters of luminosity and temperature at fixed mass ($\rho_{tl|m}=0.53\pm0.10$).

\citet{man+al16} studied the thermodynamic quantities of 40 massive clusters with $T \ga 5~\text{keV}$ identified as being dynamically relaxed and hot. Being the cluster relaxed, they identified the hydrostatic mass with the true mass and measured the off-diagonal terms of the scatter covariance matrix. They considered gas mass, core excised temperature, and core-excised or  core-included soft-band [0.1-2.4]~keV luminosity within $r_{500}$. They found $\rho_{tl|m}=-0.06\pm0.24$ and $\rho_{tm_\text{g}|m}=-0.18\pm0.28$, consistent with zero, and positive correlation between the core-included luminosity and the gas mass, $\rho_{lm_\text{g}|m}=0.43\pm0.22$. The correlation is even stronger considering core-excised luminosity, $\rho_{l_\text{ce}m_\text{g}|m}=0.88\pm0.06$.

Based on the analysis of 12 LoCuSS (Local Cluster Substructure Survey) massive clusters with $T \ga 5~\text{keV}$, \citet{oka+al10b} derived a 68.3 per cent confidence lower limit of $\rho_{tm_\text{g}|m}=0.185$, suggesting positive correlation between temperature and gas mass.

We caution that slope and normalization of the scaling relations studied here cannot be straightly compared to most literature results for two main reasons. First, we consider quantities within fixed physical radii rather than within $r_{500}$. 

Secondly, we are interested in model variables and we consider scatter in both the $X$ and the $Y$ variables. The luminosity-temperature relation determined e.g. in \citetalias{xxl_III_gil+al16} is the relation between the intrinsic luminosity and temperature and the related scatter measures the dispersions of the luminosities at a given temperature. The $l$-$t$ studied here is the hypothetical relation we would measure if temperature and luminosity were unscattered. This is the relation to study if we want to find the less scattered proxy since scatters are measured with respect to a basic third property. Similar considerations apply to the gas mass-temperature relation and \citetalias{xxl_XIII_eck+al16}. 

Notwithstanding the previous caveats, our results compare well with other recent studies. The $l$-$t$ and the $m_\text{g}$-$t$ relations for the XXL-100-GC sample were analyzed by \citetalias{xxl_III_gil+al16} and \citetalias{xxl_XIII_eck+al16}, which we refer to for detailed analysis and review of literature results. Taking into account selection effects, \citetalias{xxl_III_gil+al16} found a bolometric luminosity-temperature relation steeper than the self-similar expectation. \citetalias{xxl_XIII_eck+al16} found a gas mass-temperature relation steeper than the self-similar expectation, with a slope in agreement with \citet{arn+al07}, who analyzed ten nearby relaxed clusters. \citet{lov+al15} analyzed \emph{XMM-Newton} observations for a complete sample of local ($z<0.034$), flux-limited galaxy groups selected from the ROSAT All-Sky. They found a steeper than self-similar luminosity-temperature relation and a luminosity-gas mass relation compatible with expectations. \citet{ket+al15} investigated groups from the XMM-CFHTLS survey together with high-mass systems from the Canadian Cluster Comparison Project and low-mass systems from the Cosmic Evolution Survey to find a steep luminosity-temperature relation.

\section{Conclusions}
\label{sec_conc}

Advancements in statistical methods applied to cluster physics have made it possible the detailed study of scaling relations. The CoMaLit approach exploited in the present paper shares important features with other Bayesian methods, e.g. \citet{kel07}, \citet{oka+al10b}, \citet{roz+al10}, \citet{evr+al14}, \citet{an+co14}, \citet{mau14}, and \citet{man+al15}. The common theory behind these approaches comprises: the distinction of measured values, intrinsic scattered values, and model values; the modeling of the scaling relations as conditional probabilities; the modeling of the completeness function of the sample in terms of the intrinsic distributions of the underlying quantities. Some important differences can arise from: the treatment of the time-evolution of scaling and scatter; the treatment of uncertainties, scatters, and covariances; the modeling of the distribution of the covariate variable; the treatment of selection biases; the adopted priors.

Our analysis required neither the knowledge of the mass nor the assumption of hydrostatic equilibrium nor spherical symmetry. We only assumed that the matter halos that host clusters of galaxies have X-ray observable properties that are log-normally distributed around power-law scaling relations in halo mass. 

We may think of the basic cluster property $Z$ as a `mass substitute', i.e. the best statistical quantity to use when we have no access to total mass. It should be the quantity with the strongest correlation to the studied properties. In principle, this ideal quantity may be fictitious. Only direct measurements of the cluster masses can eliminate any ambiguity or doubt. However, theoretical considerations based on the self-similar model, numerical simulations, and analyses of cluster samples with measured mass support the simplest hypothesis that $Z$ is the mass.


We studied the scaling relations between X-ray properties of the XXL-100-GC sample and the intrinsic conditional scatters without any external mass calibration. In particular, we considered the spectroscopic temperature, the soft band luminosity, and the gas mass within fixed physical radii. The sample spans one order of magnitude in temperature from small groups at $T\sim~0.6~\text{keV}$ to more massive clusters at $T\sim7~\text{keV}$. This probes the lower end of the halo mass function at the group scale, whereas most of previous analyses focused on more massive clusters ($T \ga 4~\text{keV}$). The gas mass confirms itself as an excellent proxy. Even when measured within a fixed physical length, cluster properties can be recovered from gas masses with $\sim 8$ per cent accuracy. 

Noteworthily, the gas mass is an equal or even better proxy to the weak lensing mass, which has an intrinsic scatter of $\sigma_\text{WL} \sim$ 15 per cent (\citetalias{se+et15_comalit_I}; \citealt{man+al15}) and a better proxy than the hydrostatic mass, $\sigma_\text{HE} \sim$ 25 per cent \citepalias{se+et15_comalit_I}. 

We considered only a subsample of X-ray properties. Other proposals as mass proxies can be appealing too. The integrated SZ Compton parameter $Y_\text{SZ}$ is expected to be tightly correlated to the energy content and the total mass of the clusters \citep{se+et15_comalit_IV}, but its measurements can be elusive for small systems. The product of the temperature and $M_\text{gas,500}$, $Y_\text{X}$, is viewed as a robust mass indicator with low-scatter \citep{kra+al06}. However, this proxy best performs if the temperature is core-excised, which is not practical in small groups. Furthermore, any positive correlation between intrinsic scatters of temperature and gas mass, as found in this paper, can worsen its performance. 

Multi-probe analyses can open new windows on the evolution and formations of structures \citep{ser+al18_CLUMP_I}. Halo properties can be better understood in terms of multi-dimensional plans than basic one-to-one relations \citep{fuj+al18b}. Generalized scaling laws suitably weighting X-ray observables have to be considered to calibrate the proxy with the minimum scatter \citep{ett13}.

We retrieved positive correlation between measured gas mass, temperature, and luminosity. The study of covariance between intrinsic scatters is important as it impacts the propagation of selection biases based on one observable to biases on other observable quantities \citep{mau14}. For example, without taking the covariance between luminosity and gas mass or temperature into account, cluster masses estimated from $M_\text{g}$ or $T$ in an X-ray flux-limited sample would be biased high, with implications for cosmological studies \citep{mau14}.

The simple assumption of underlying power-law relations is enough to estimate the intrinsic scatters of the observed properties and rank them. However, the loop cannot be closed without the mass information.  This is needed to calibrate the scaling relation and confirm that the optimal cluster proxy is indeed the optimal mass proxy. 



\section*{Acknowledgements}

XXL is an international project based around an XMM Very Large Programme surveying two 25 deg$^2$ extragalactic fields at a depth of $\sim5\times 10^{-15}$erg cm$^{-2}$ s$^{-1}$ in the [0.5-2] keV band for point-like sources. The XXL website is \url{http://irfu.cea.fr/xxl/}. Multi-band information and spectroscopic follow-up of the X-ray sources are obtained through a number of survey programmes, summarised at \url{http://xxlmultiwave.pbworks.com/}.

MS and SE acknowledge financial contribution from the contract ASI-INAF n.2017-14-H.0. The Saclay group acknowledges long-term support from the Centre National d'Etudes Spatiales (CNES).
This research has made use of NASA's Astrophysics Data System (ADS) and of the NASA/IPAC Extragalactic Database (NED), which is operated by the Jet Propulsion Laboratory, California Institute of Technology, under contract with the National Aeronautics and Space Administration.


\appendix

\section{Asymmetric errors}
\label{app_asym}

The likelihood function of the measured spectroscopic temperature is approximately Gaussian in log space \citepalias{xxl_III_gil+al16}. We used the method of \citet{and12} to convert the asymmetric errors on $T$ computed by XSPEC to symmetric errors on $\log T$. If the probability distribution $p(T)$ is approximately log-normal, the standard deviation of the distribution $p(\log T)$ is given by 
\beq
\delta_{\log T}\simeq  \frac{1}{2}  \log \left( \frac{T_\text{mode} + \delta^+}{T_\text{mode} - \delta^-} \right)
\eeq
where $T_\text{mode}$ is the mode and the uncertainties $\delta^+$ and $\delta^-$ are the points where the likelihood is lower than its maximum by a factor $\exp(-1/2)$. 

Under the same assumption, we extended the prescription of \citet{and12} and we also computed the mean as 
\beq
\mu_{\log T} \simeq    \log (T_\text{mode}) + \delta_{\log T}^2\ln(10) .
\eeq

\section{2D  posteriors}
\label{app_2D}

\begin{figure*}
\resizebox{\hsize}{!}{\includegraphics{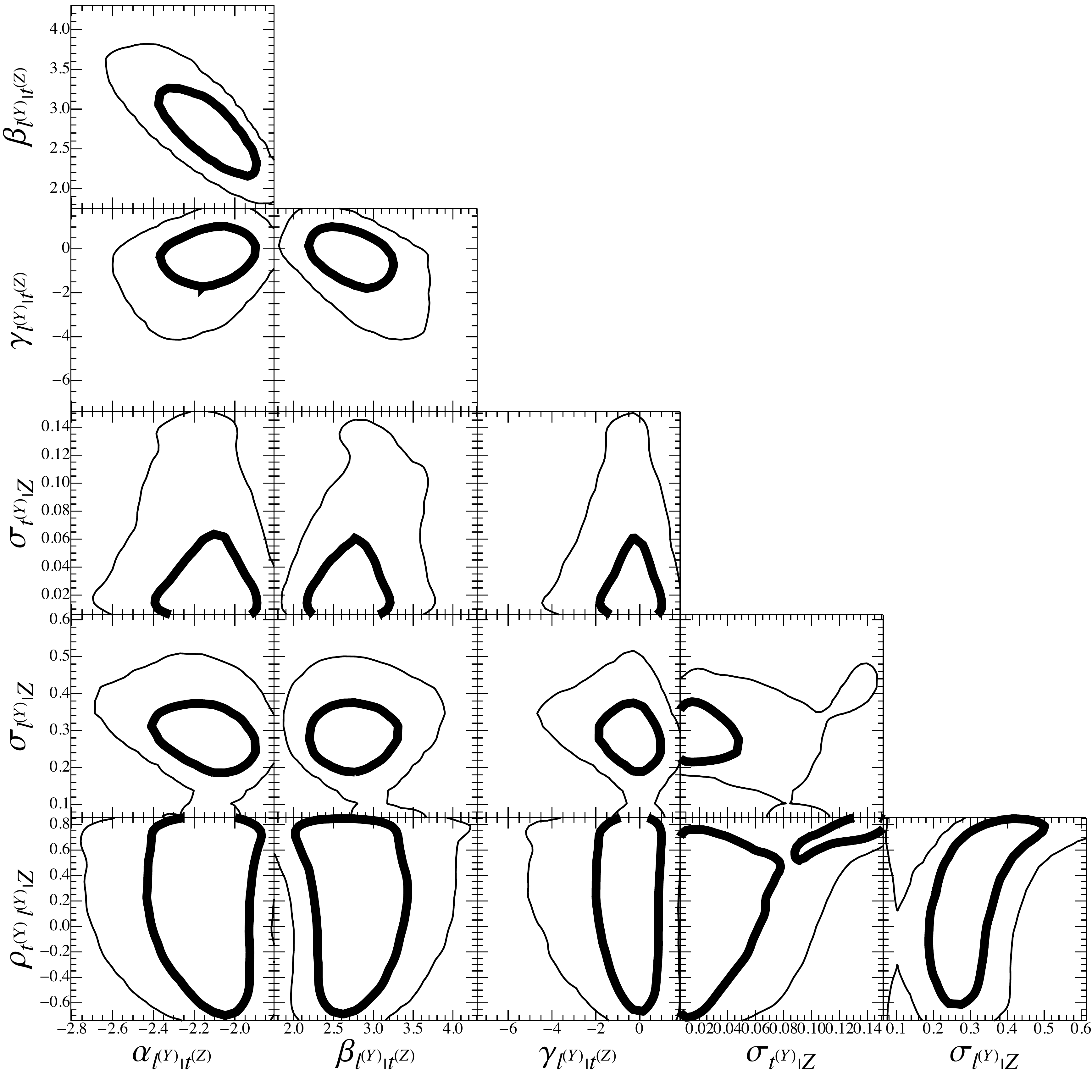}}
\caption{Probability distributions of the parameters of the scaling relation between luminosity and temperature, $l$-$t$. The thick (thin) lines include the 1-(2-)$\sigma$ confidence region in two dimensions, here defined as the region within which the probability is larger than $\exp[-2.3/2]$ ($\exp[-6.17/2]$) of the maximum.
}
\label{fig_lx_tx_PDF_2D}
\end{figure*}

\begin{figure*}
\resizebox{\hsize}{!}{\includegraphics{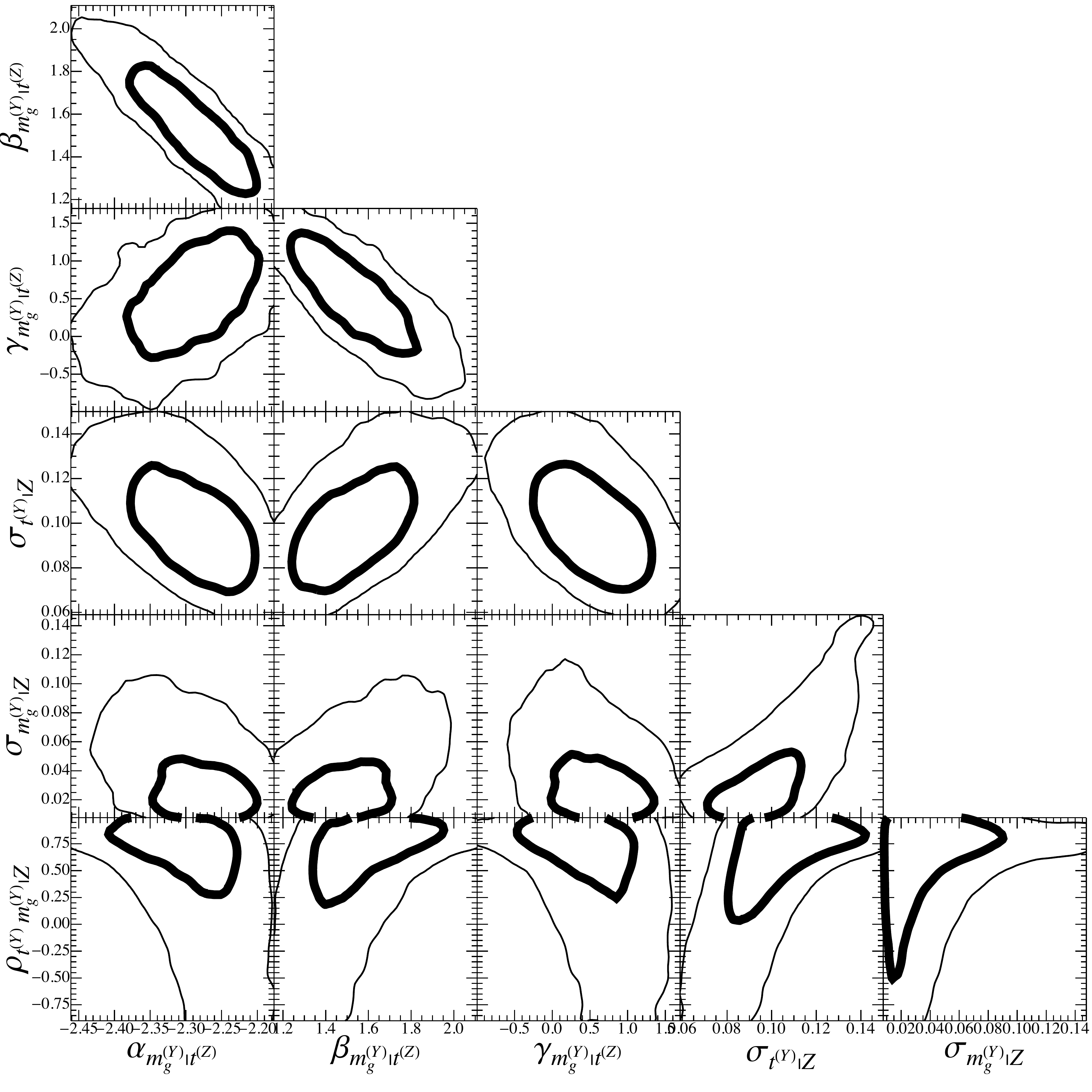}}
\caption{Probability distributions of the parameters of the scaling relation between gas mass and temperature, $m_\text{g}$-$t$. The thick (thin) contours include the 1-(2-)$\sigma$ confidence region in two dimensions, here defined as the region within which the probability is larger than $\exp[-2.3/2]$ ($\exp[-6.17/2]$) of the maximum.}
\label{fig_mg_tx_PDF_2D}
\end{figure*}

\begin{figure*}
\resizebox{\hsize}{!}{\includegraphics{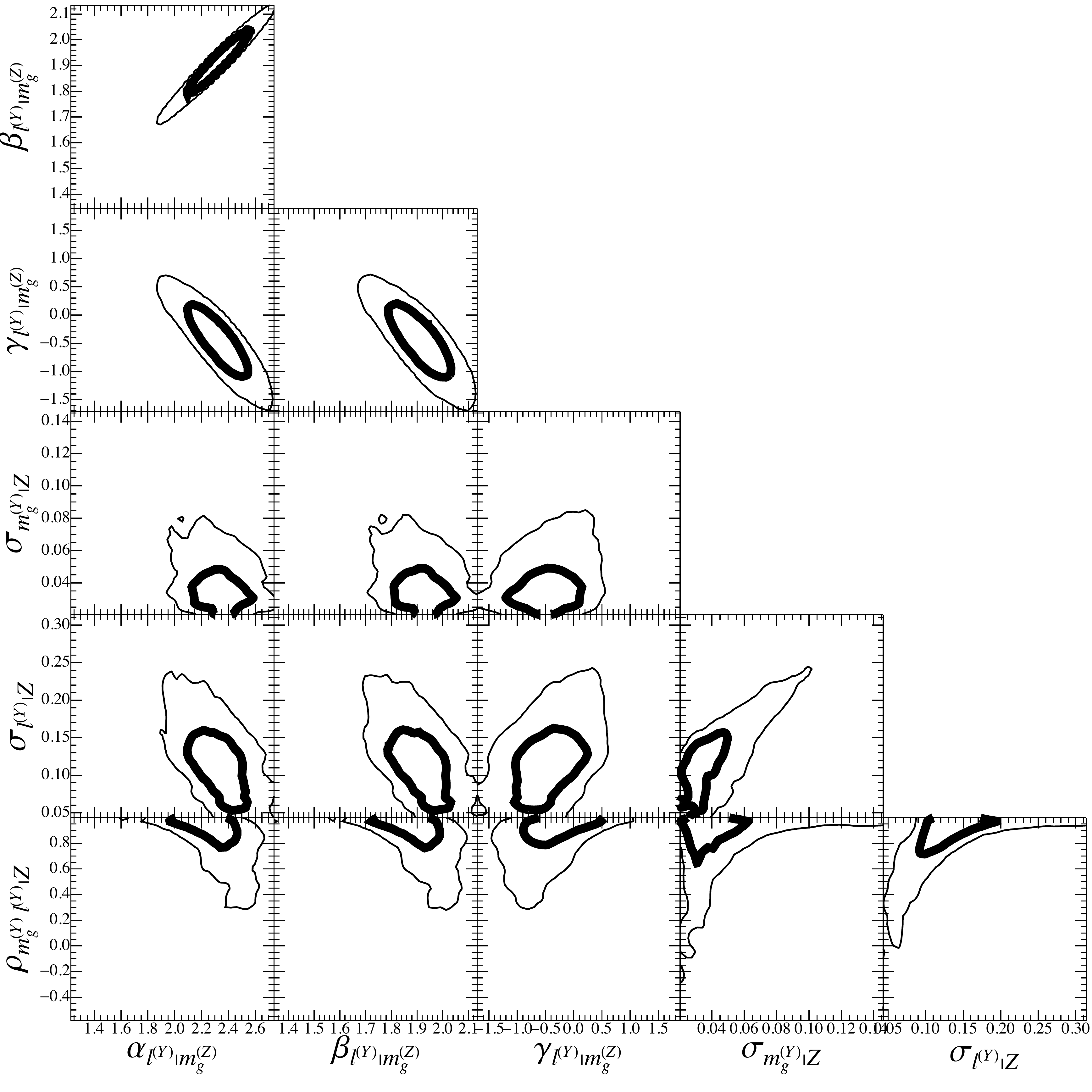}}
\caption{Probability distributions of the parameters of the scaling relation between luminosity and gas mass, $l$-$m_\text{g}$. The thick (thin)  contours include the 1-(2-)$\sigma$ confidence region in two dimensions, here defined as the region within which the probability is larger than $\exp[-2.3/2]$ ($\exp[-6.17/2]$) of the maximum.
}
\label{fig_lx_mg_PDF_2D}
\end{figure*}

\begin{figure*}
\resizebox{\hsize}{!}{\includegraphics{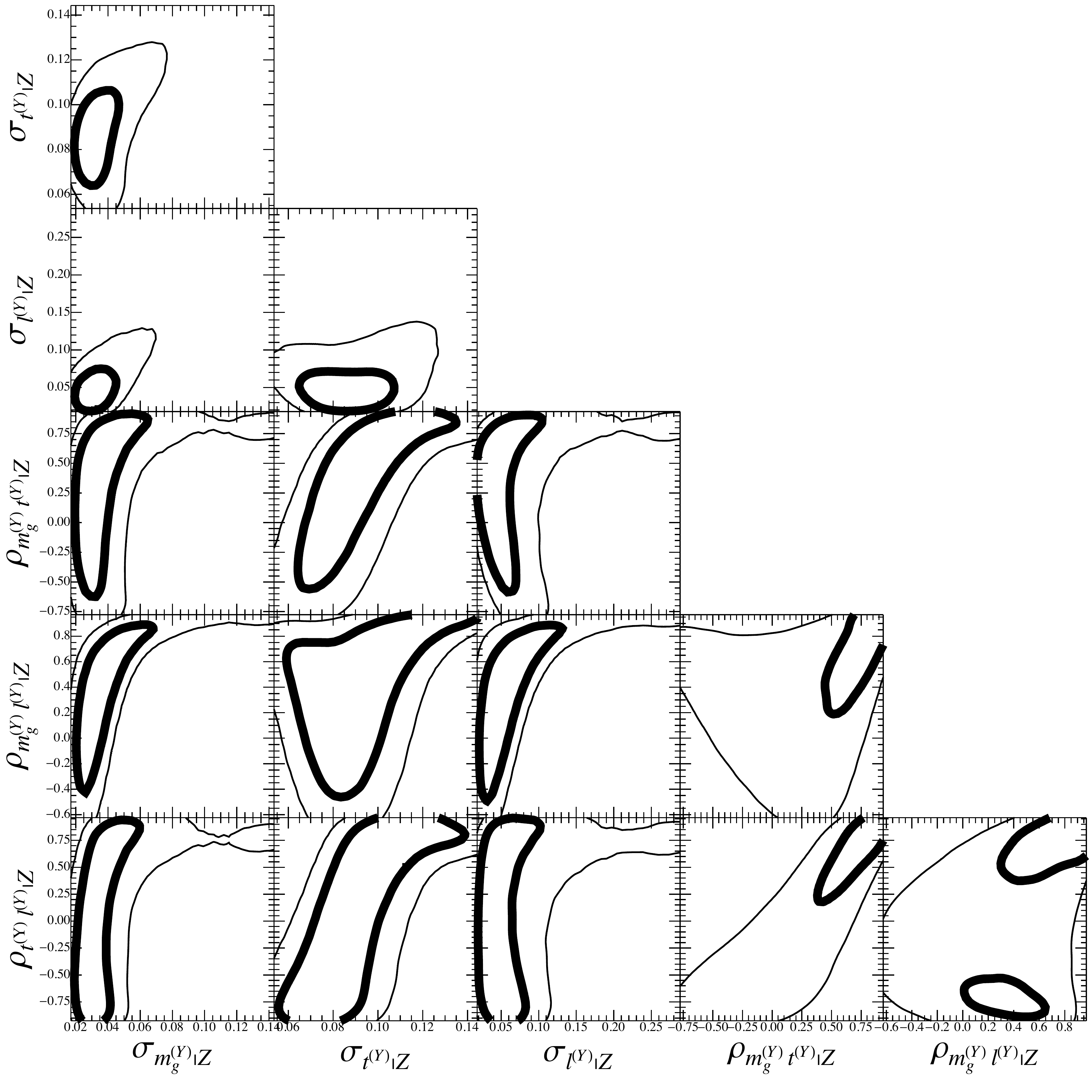}}
\caption{Probability distributions of the scatter parameters from the multi-response regression. The thick (thin)  contours include the 1-(2-)$\sigma$ confidence region in two dimensions, here defined as the region within which the probability is larger than $\exp[-2.3/2]$ ($\exp[-6.17/2]$) of the maximum.}
\label{fig_mg_tx_lx_PDF_2D}
\end{figure*}

The marginalized 2D posteriori probabilities for the luminosity-temperature, gas mass-temperature, and luminosity-gas mass relations are plotted in Figs.~\ref{fig_lx_tx_PDF_2D}, \ref{fig_mg_tx_PDF_2D}, and \ref{fig_lx_mg_PDF_2D}. The results for the multi-response regression are shown in Fig.~\ref{fig_mg_tx_lx_PDF_2D}.

\end{document}